\documentclass[a4paper,fleqn,usenatbib]{mnras}

\usepackage{newtxtext,newtxmath}
\usepackage[T1]{fontenc}
\usepackage{ae,aecompl}

\usepackage{graphicx}	
\usepackage{amsmath}	
\usepackage{amssymb}	
\usepackage[compatibility=false]{caption}
\usepackage{subcaption}

\newcommand{\mi}{\relax \ifmmode {\mu{\mbox m}}\else $\mu$m\fi}
\newcommand{\sii}{\relax \ifmmode {\mbox S\,{\scshape ii}}\else S\,{\scshape ii}\fi}
\newcommand{\siii}{\relax \ifmmode {\mbox S\,{\textsc {iii}}}\else S\,{\scshape iii}\fi}
\newcommand{\siv}{\relax \ifmmode {\mbox S\,{\textsc {iv}}}\else S\,{\scshape iv}\fi}
\newcommand{\nii}{\relax \ifmmode {\mbox N\,{\scshape ii}}\else N\,{\scshape ii}\fi}
\newcommand{\neii}{\relax \ifmmode {\mbox Ne\,{\textsc {ii}}}\else Ne\,{\scshape ii}\fi}
\newcommand{\neiii}{\relax \ifmmode {\mbox Ne\,{\textsc {iii}}}\else Ne\,{\scshape iii}\fi}
\newcommand{\oiii}{\relax \ifmmode {\mbox O\,{\scshape iii}}\else O\,{\scshape iii}\fi}
\newcommand{\oii}{\relax \ifmmode {\mbox O\,{\scshape ii}}\else O\,{\scshape ii}\fi}
\newcommand{\oi}{\relax \ifmmode {\mbox O\,{\scshape i}}\else O\,{\scshape i}\fi}
\newcommand{\ha}{\relax \ifmmode {\mbox H}\alpha\else H$\alpha$\fi}
\newcommand{\hep}{\relax \ifmmode {\mbox H}\epsilon\else H$\epsilon$\fi}
\newcommand{\hdel}{\relax \ifmmode {\mbox H}\delta\else H$\delta$\fi}
\newcommand{\hgam}{\relax \ifmmode {\mbox H}\gamma\else H$\gamma$\fi}
\newcommand{\doclogO}{\relax \ifmmode {\mbox 12 + log(O/H)}\else  12 + log(O/H)}

\newcommand{\pa}{\relax \ifmmode {\mbox Pa}\alpha\else Pa$\alpha$\fi}
\newcommand{\hb}{\relax \ifmmode {\mbox H}\beta\else H$\beta$\fi}
\newcommand{\rdostres}{\relax \ifmmode {\,\mbox{R}}_{\rm 23}\else \,\mbox{R}$_{\rm 23}$\fi}
\newcommand{\ergs}{\relax \ifmmode {\,\mbox{erg\,s}}^{-1}\else \,\mbox{erg\,s}$^{-1}$\fi}
\newcommand{\me}{\relax \ifmmode {\,}^{-1}\else \,$^{-1}$\fi}
\newcommand{\degree}{$^\circ$}

\newcommand{\msun}{\relax \ifmmode {\,\mbox{M}}_{\odot}\else \,\mbox{M}$_{\odot}$\fi}

\newcommand{\cmtres}{\relax \ifmmode {\,\mbox{cm}}^{-3}\else \,\mbox{cm}$^{-3}$\fi}
\newcommand{\cmdos}{\relax \ifmmode {\,\mbox{cm}}^{-2}\else \,\mbox{cm}$^{-2}$\fi}
\newcommand{\cmseis}{\relax \ifmmode {\,\mbox{cm}}^{-6}\else \,\mbox{cm}$^{-6}$\fi}
\newcommand{\hi}{\relax \ifmmode {\mbox H\,{\scshape i}}\else H\,{\scshape i}\fi}
\newcommand{\arcminut}{$^{\prime}$}

\newcommand{\Spi}{{\it Spitzer}}
\newcommand{\Her}{{\it Herschel}}

\newcommand{\chidos}{$\chi^{2}_{\rm red}$}
\newcommand{\hii}{\relax \ifmmode {\mbox H\,{\scshape ii}}\else H\,{\scshape ii}\fi}

\title[Metals and dust in M\,101 and NGC\,628]{Metals and dust content across the galaxies M\,101 and NGC\,628}
\author[J.M. V\'{\i}lchez et al.]{J.M. V\'{\i}lchez$^{1}$\thanks{E-mail: jvm@iaa.es, Visiting fellow, Institute of Astronomy, University of Cambridge},
M. Rela\~no$^{2,3,4}$,
R. Kennicutt$^{2,5}$,
I. De Looze$^{6,7}$,
M. Moll{\'a}$^{8}$, 
M. Galametz$^{9}$
\\
$^{1}$Instituto de Astrof\'{\i}sica de Andaluc\'{\i}a - CSIC, Glorieta de la Astronom\'{\i}a s.n., 18008 Granada, Spain\\
$^{2}$Institute of Astronomy, University of Cambridge, Madingley Road, Cambridge CB3 0HA, UK\\
$^{3}$Dept. F\'{i}sica Te\'orica y del Cosmos, Universidad de Granada, Spain\\
$^{4}$Instituto Universitario Carlos I de F\'isica Te\'orica y Computacional, Universidad de Granada, 18071, Granada, Spain\\
$^{5}$Steward Observatory, University of Arizona, 933 North Cherry Avenue, Tucson, AZ 85721\\
$^{6}$Department of Physics and Astronomy, University College London, Gower Street, London WC1E 6BT, UK\\
$^{7}$Sterrenkundig Observatorium, Universiteit Gent, Krijgslaan 281 S9, B-9000 Gent, Belgium\\
$^{8}$Departamento de Investigaci\'on B\'asica, CIEMAT, E-28040 Madrid, Spain\\
$^{9}$Astrophysics department, CEA/DRF/IRFU/DAp, Université Paris Saclay, UMR AIM, F-91191 Gif-sur-Yvette, France
}

\date{Accepted XXX. Received YYY; in original form ZZZ}

\pubyear{2018}

\begin{document}
\label{firstpage}
\pagerange{\pageref{firstpage}--\pageref{lastpage}}
\maketitle

\begin{abstract}
We present a spatially resolved study of the relation between dust and metallicity in the nearby spiral galaxies M\,101 (NGC\,5457)  and NGC\,628 (M\,74). We explore the relation between the chemical abundances of their gas and stars with their dust content and their chemical evolution. The empirical spatially resolved oxygen effective yield and the gas to dust mass ratio (GDR) across both disc galaxies are derived, sampling one dex in oxygen abundance. We find that the metal budget of the NGC\,628 disc and most of the M\,101 disc appears consistent with the predictions of the simple model of chemical evolution for an oxygen yield between half and one solar, whereas the outermost region (R\,$\geq$\,0.8\,$\rm R_{25}$) of M\,101 presents deviations suggesting the presence of gas flows. The GDR-metallicity relation shows a two slopes behaviour, with a break at 12+log(O/H)\,$\approx$\,8.4, a critical metallicity predicted by theoretical dust models when stardust production equals grain growth. A relation between GDR and the fraction of molecular to total gas, $\rm\Sigma_{H_{2}}$/$\rm\Sigma_{gas}$ is also found. We suggest an empirical relationship between GDR and the combination of 12+log(O/H), for metallicity, and $\rm\Sigma_{H_{2}}$/$\rm\Sigma_{gas}$, a proxy for the molecular clouds fraction. The GDR is closely related with metallicity at low abundance and with $\rm\Sigma_{H_{2}}$/$\rm\Sigma_{gas}$ for higher metallicities suggesting ISM dust growth. The ratio $\rm\Sigma_{dust}$/$\rm\Sigma_{star}$ correlates well with 12 + log(O/H) and strongly with log(N/O) in both galaxies. For abundances below the critical one, the 'stardust' production gives us a constant value suggesting a stellar dust yield similar to the oxygen yield. 

\end{abstract}


\section{Introduction}
The study of the chemical evolution of galaxies is a powerful way to understand of their history of star formation and evolution. Nucleosynthesis of the different chemical elements occurs in stars of different masses and lifetimes that inject their metal production into the interstellar medium (ISM) across the galaxies. Chemical abundances of stars and ISM in star forming disc galaxies are observed to be different depending on the precise location in the galaxy, most frequently defining spatial gradients that reflect its star formation and chemical history. Massive gas flows have proven to be an essential ingredient in order to explain the observed shapes of the abundance gradients in disc galaxies. In particular outflows, and especially selective enriched outflows, were postulated for virtually all galaxies; mainly because they seem to be an efficient mechanism to decrease the yield of the chemical system \citep[e.g.][]{2007ApJ...658..941D}, though some discussion is still open \citep[e.g.][]{2001ApJ...552...91S,2013MNRAS.434.2491G}. On the other hand gas inflows are a key ingredient in most chemical evolution codes as well as in accretion models of galaxy formation.

There is an important amount of work in the literature devoted to study the (inter)relation between the properties of disc galaxies, notably spirals, and their chemical evolution parameters. Large part of this work is based on galaxy global properties (e.g. from SDSS), i.e. under the {\it one point-one galaxy} scheme. Spatially resolved chemical evolution studies can provide new useful information. Studying the chemical evolution of discs at spatially resolved scales imposes strong constraints (e.g. to selective metal outflows) linked to galaxy structure and dynamics. For example, supernova driven galactic winds \citep[e.g.][]{1974MNRAS.169..229L,1983A&A...123..121M} departing from closed-box evolution, would only produce substantial decrease of the retaining fraction of processed gas for galaxies with rotation velocities lower than some 160 km/s \citep[e.g.][]{2000ApJS..129..493H,2004A&A...425..849P,2004ApJ...613..898T}. 

A useful empirical indicator of the efficiency of the enrichment of the interstellar gas in heavy elements is the "effective yield" \citep{1990MNRAS.246..678E,2007ApJ...658..941D}, a quantity that measures how much the metallicity of a galaxy deviates from what would be expected for a system with the same gas mass fraction, but that have evolved in a closed box framework, i.e. where flows of gas (inflow or outflow) are not allowed (see Sect.~\ref{sec:chemevol}). 
As shown by \citet{2007ApJ...658..941D}, in fact only metal-enriched outflows can reduce the effective yield of gas-rich systems by a substantial amount. For gas-poor systems, chemical yields are difficult to be changed irrespective of the amount of gas lost or accreted, though. 

The study of outflows and inflows of gas are important to understand the chemical evolution of galaxies since both processes are directly related to the large reservoirs of neutral and molecular gas and also circumgalactic material embracing star forming galaxies \citep[e.g.][]{2017ARA&A..55..389T}. This external reservoirs can be metal enriched 
from supernova winds leaving chemically processed gas. Models can include also external gas infall, either pristine or enriched. Though frequently overlooked, incorporating spatially resolved chemical abundances, e.g. O/H radial gradients, can add new constraints to chemical evolution modelling. Also, spatially resolved information on the ratio of secondary to primary elements (e.g. nitrogen to oxygen ratio, N/O, versus O/H) can tell us on the relative roles played by flows, the star formation efficiency, or possible delays in the delivery of the nucleosynthesis of low and intermediate mass stars, among other clues \citep[e.g.][]{2006ApJ...647..984H,2006MNRAS.372.1069M, 2017MNRAS.471.1743D, 2018MNRAS.473..241H}.    

Much of the work done for samples of galaxies has used single aperture abundances \citep[e.g.][]{1994ApJ...420...87Z,2002ApJ...581.1019G,2004ApJ...613..898T} losing the wealth of information across e.g. spiral discs, whereas recent integral field spectroscopy surveys (e.g. CALIFA; MANGA, SAMI; VENGA) have provided new spatially resolved data. Measuring chemical abundance for many positions across galaxies -from center to outskirts- allows a better characterisation of the metallicity spatial profile along the discs that, together with the gas mass fraction profile, give strong constraints on the effective yield and chemical modelling. The abundance profiles of O/H and N/O, taken together with their corresponding gas and stellar surface density, can inform us also on how and when gas inflows and/or outflows are present, modulating the chemical, ionising and mechanical feedback of each star forming complex on the molecular clouds and dust reservoirs. These feedback mechanisms could affect the places where dust growth is believed to be produced, eventually leading to e.g. grain destruction from supernova blast waves and (possibly also) winds shocks or intense radiation fields.  

Together with metallicity and gas, dust is also directly related, via its formation and destruction channels, to the chemical evolution of galaxies. Dust constitutes a fundamental ingredient for star formation and chemical evolution often overlooked in the metal balance across galaxies. Dust is an essential component of the ISM; absorbs and scatter the light from stars and reemits in the infrared dominating the spectral energy distribution in this range. Dust grains are intimately related to the formation of molecular hydrogen, both providing substantial cooling, of especial relevance at very low metallicity \citep[e.g.][]{2005ApJ...626..627O,2006MNRAS.369.1437S}.

A detailed description of the physics and the chemical properties of dust, grain distribution and composition in the ISM is out of the scope of this work and can be found elsewhere \citep[e.g.][]{2013EP&S...65..213A,2013MNRAS.432..637A,2014A&A...563A..31R, 2017MNRAS.467..699H,2017arXiv171107434G}. Dust production and destruction and dust content evolution involve very complex physics and different chemical processes competing, in practice, through diverse channels which, in turns, modulate the final dust to gas mass ratio (GDR) \citep[e.g.][]{2013EP&S...65..213A,2014A&A...562A..76Z,2017arXiv171107434G}. Production by evolved stars and supernovae adds to the dust budget, and the overall dust yield from stellar sources (basically AGB stars and some SNe II) can be evaluated for a given IMF and star formation rate history. Nonetheless, stellar related dust production alone does not support current observations, and growth of grains in the ISM is needed especially in high metallicity environments.

Models of the dust content in galaxies \citep[e.g.][]{2013EP&S...65..213A} show how an increasing rate of dust mass growth in the ISM can dominate dust production by stars assuming constant star formation. Also \citet{2014A&A...562A..76Z} has included the effects of episodic star formation histories on the ISM dust growth. On the other hand, destruction of dust can be easily produced due to supernova shocks \citep{2007MNRAS.378..973B}, and also intense UV radiation fields have been proposed to affect to the carbonaceous dust population \citep{2017MNRAS.471.1743D}.  A critical metallicity has been defined by \citet{2013EP&S...65..213A}  which discriminates between these two main ISM dust regimes which should be seen when analysing the GDR as a function of the gas oxygen abundance, assuming different star formation time scales or a bursty star formation scheme \citep[e.g.][]{2014A&A...562A..76Z}.  \citet{2014A&A...563A..31R} studied the GDR versus oxygen abundance relation for an extended sample of star forming galaxies, including from dwarfs to large spirals. Using models from \citet{2014A&A...562A..76Z} and \citet{2013EP&S...65..213A}, they showed the scatter they obtain in this relation could be explained for a range of chemical evolution models and star formation timescales and suggested two main trends with metallicity.
It is important to bear in mind that the derivation of the GDR is complex and can suffer from several effects; among them the conversion of CO observations into molecular gas content is a large factor of uncertainty, and the impact of the CO-dark gas fraction, as well as the dust model or the CO-chemical abundance gradients; i.e. the lower oxygen abundance favouring the prominence of more atomic gas and more CO-dark gas, probably due to a lower shielding of dust and self-shielding, as suggested for the lower metallicity outer Milky Way \citep[e.g.][]{2017A&A...606L..12G}. Recent studies of spatially resolved galaxies \citep[e.g.][]{2013ApJ...777....5S} have tried to minimise the uncertainty in the CO-H$_{\rm 2}$ conversion factor on kpc scales. \citet{2017A&A...605A..18C} have presented a new database of gas and dust profiles for a large sample of galaxies, proposing another recipe to tackle this problem. 

It is clear that the star formation history, metal content and chemical evolution across a star forming galaxy should appear tightly related to the observed dust content.
A detailed study relating the dust content and metallicity in galaxy discs using high quality data is timely. We need to combine chemical abundances and physical conditions of the ISM with a precise determination of the dust content and evolution. We have selected the nearly face on nearby galaxies NGC\,5457 (M\,101) and NGC\,628 (M\,74) as prototype objects to carry out this study. They have a similar mass (log\,M$_\star$/M$_{\odot}$\,$\approx$\,10.2) though very different chemical and structural properties and environment. The global budget of metals and dust across M\,101 and NGC\,628 are derived from various datasets covering a large range of diagnostics (optical spectroscopy of the ionised gas, multi-band spectrophotometry of the stars, dust and gas, radio and mm observations of their neutral and molecular gas) analysed consistently. The GDR and the empirical spatially resolved (effective) metal yield across the discs of both galaxies are derived, sampling one dex range in O/H. Here we empirically explore the presence of a relationship between GDR and chemical abundance across both galaxies, discussing the effects of their different chemical evolution, from their effective yield spatial profiles to the dust yield and ISM dust growth.

This paper is organised as follows: in Sect.~\ref{sec:data} we describe the data sets used and provide a description of the two selected galaxies. The main results obtained in this work are detailed in Sect.~\ref{sec:results}. The final discussion, Sect.~\ref{sec:discussion}, and main conclusions, Sect.~\ref{sec:conclusions}, are presented.

\section{Sample and data}\label{sec:data}

\begin{figure*} 
\includegraphics[width=0.45\textwidth]{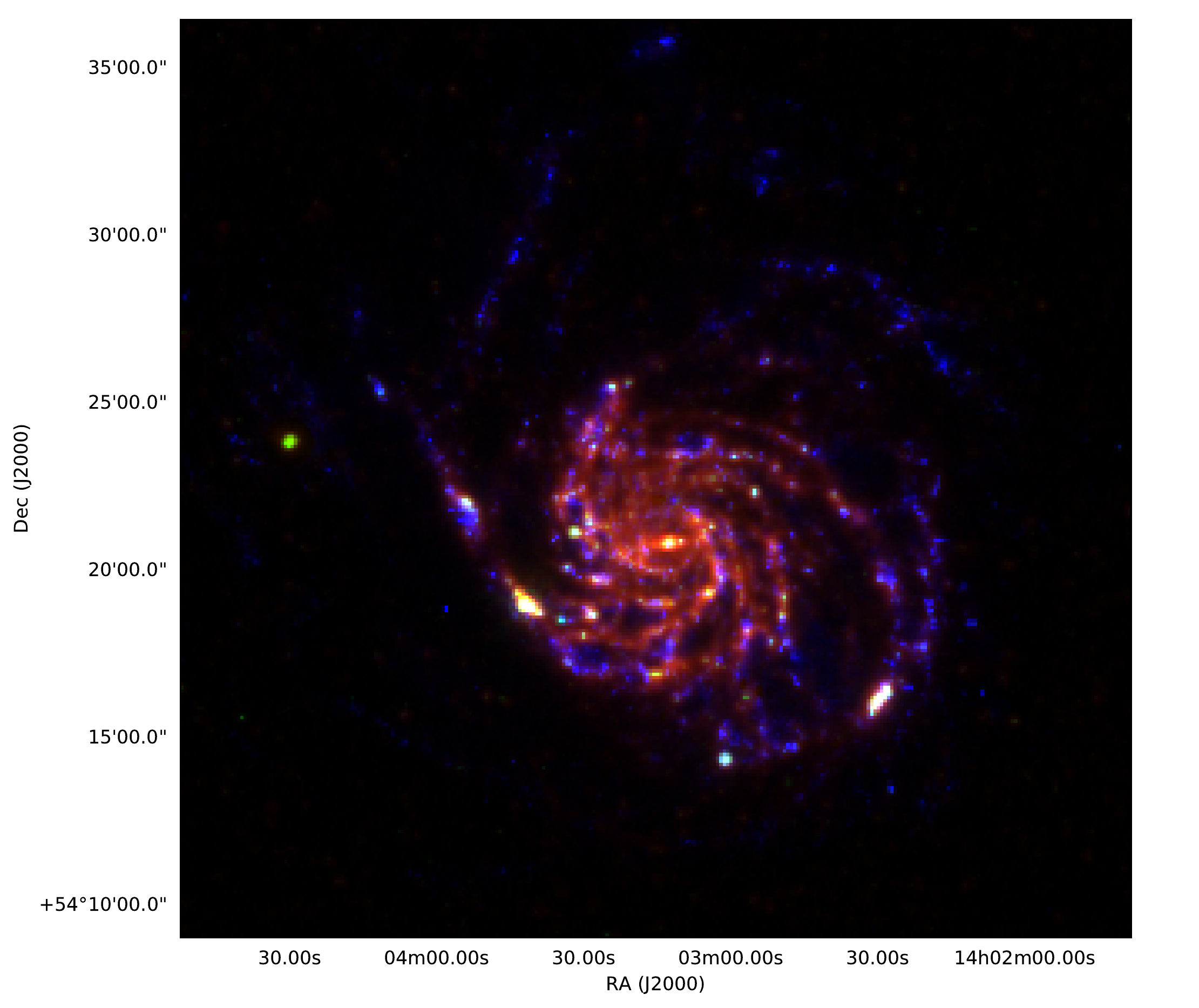}
\includegraphics[width=0.45\textwidth]{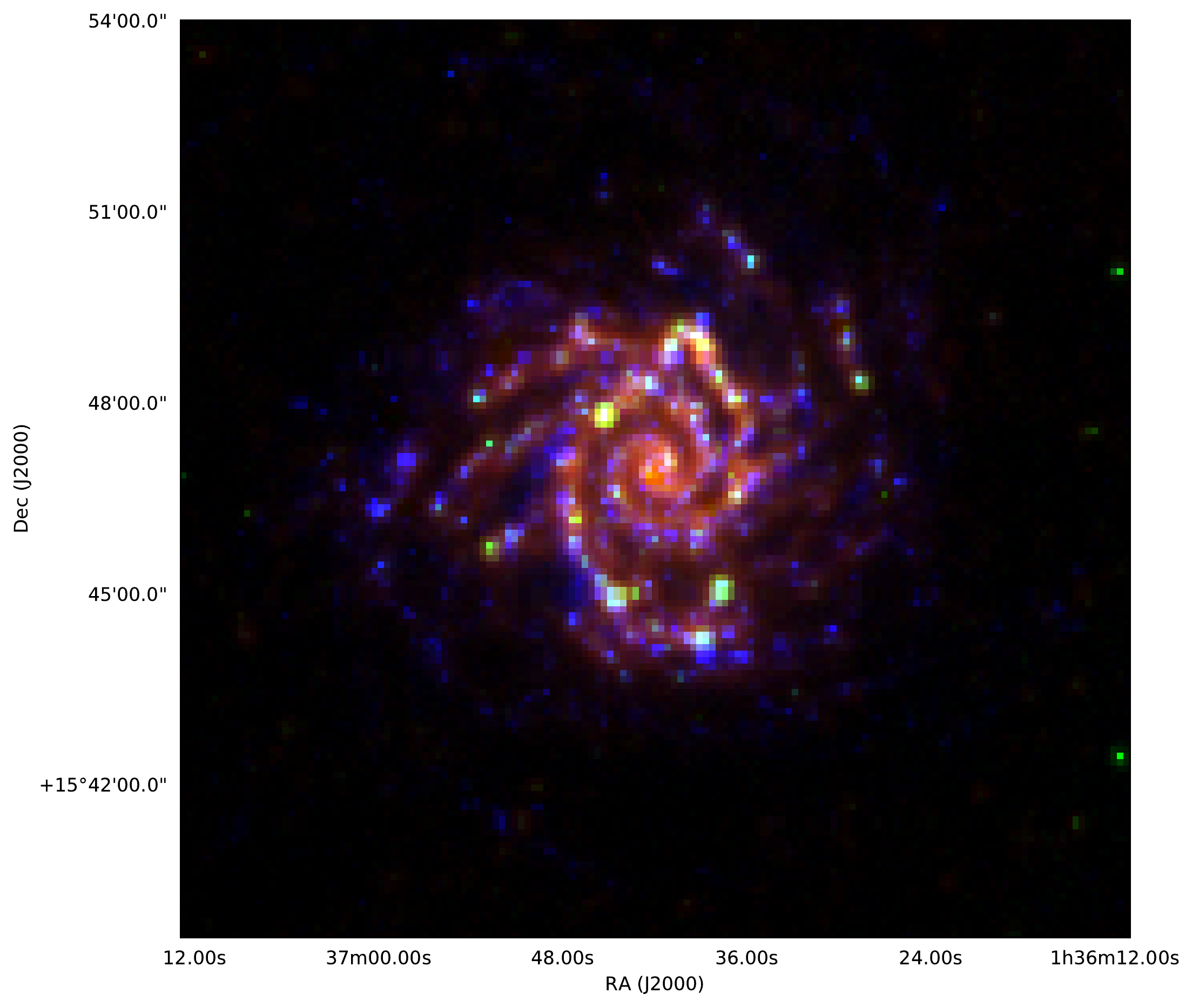}
   \caption{RGB images for  M\,101 (left)  and NGC\,628 (right). GALEX FUV image in blue, 24\,\mi\ in green and 250\,\mi\ in red, trace the stellar continuum, hot and cool dust emission, respectively.}
   \label{fig:RGB}
\end{figure*}

The galaxies M\,101 (NGC\,5457) and NGC\,628 (M\,74), are the two selected nearby well resolved spirals taken as representative examples of their class. 
M\,101 is a nearly face on (inclination 18\degree) large spiral located at 7.4\,Mpc distance. The galaxy shows a well-known asymmetry of the disc and an asymmetric plume towards larger radius seen in deep exposures consistent with a very extended HI disc to the northeast \citep{2013ApJ...762...82M}, a footprint of possible past interactions with the nearby galaxies NGC\,5477 and NGC\,5474. M\,101 general properties are presented in Table\,\ref{tab:1}. 
NGC\,628 (M\,74) is a giant late type spiral galaxy located at a distance of 7.2\,Mpc and seen practically face on (inclination 5\degree). The adopted properties for NGC\,628 are shown in Table\,\ref{tab:1}. NGC\,628 presents an ideally suited large extended spiral structure and an undisturbed optical profile.
M\,101 and NGC\,628  have been observed at all wavelength ranges relevant for this study, belong to the KINGFISH (Key Insights on Nearby Galaxies: a Far-Infrared Survey with Herschel) sample \citep{2011PASP..123.1347K} and high quality photometric and spectroscopic data are available for them. Finally, both galaxies have been observed as part of the CHAOS (CHemical Abundances of Spirals) project \citep{2015ApJ...806...16B,2016ApJ...830....4C} that derives estimates of the electron temperatures and direct chemical abundances for a large number of \hii\ regions. 
The structure of both galaxies has been studied within the Spitzer Survey of Stellar Structure in Galaxies (S4G) \citep{2010PASP..122.1397S} and they have been classified (surface brightness profile decomposition performed) as hosting a pseudobulge, which appears especially notable in M\,101. In Fig.~\ref{fig:RGB} we show RGB images of both galaxies.

\begin{table*}
\centering
\caption{Global properties of M\,101 (NGC\,5457) and NGC\,628 (M\,74). Units of right ascension are hours, minutes, and seconds, and units of declination are degrees, arcminutes, and arcseconds. References: [1] NASA/IPAC Extragalactic Database [2] \citet{2008AJ....136.2563W}, [3] \citet{1984A&A...132...20S}, [4] \citet{2009ApJ...697.1870E}, [5] \citet{2000ApJS..128..431F}, [6] \citet{2006PASP..118..351V}, [7] \citet{2011PASP..123.1347K}, and [8] \citet{2011MNRAS.414..538K}.}
\label{tab:1}
\begin{tabular}{lcccc}\hline
& NGC\,5457&&NGC\,628&\\
Property & Value & Reference & Value& Reference\\
\hline
R.A. & 14h03m12.5s & [1] &  01h36m41.7s & [1]    \\
Dec &  54$^\circ$20m56s & [1] & 15$^\circ$47m01s & [1]\\
Inclination  & 18$^\circ$  & [2] &  5$^\circ$  & [3]\\
Position Angle & 39$^\circ$ & [2] & 12$^\circ$  & [4]  \\
Distance & 7.4\,Mpc  & [5] & 7.2\,Mpc  & [6]\\
$\rm R_{25}$ & 864\arcsec & [7] & 315\arcsec & [8]\\
\hline
\end{tabular}
\label{tab:1}
\end{table*}

\subsection{Chemical abundances}
\subsubsection{Ionized gas} 
Chemical abundances of a large number of \hii\ regions in M\,101 have been obtained in several works, notably \citet{1998AJ....116.2805V}, \citet{2003ApJ...591..801K}, \citet{2007ApJ...656..186B} and \citet{2013ApJ...766...17L}. In this work we use only abundances derived directly using measurements of the electron temperature sensitive lines. We avoid chemical abundances derived from either empirical or photoionisation models calibrations; though a plethora of different abundance calibrations is available they can suffer from substantial uncertainties and systematics \citep[see e.g.][]{2013ApJ...766...17L}. Therefore here we have adopted the abundance gradient of oxygen and nitrogen from \citet{2016ApJ...830....4C} who derived direct abundances (i.e. using the measurements of electron temperature) for a total of 109 \hii\ regions, including updated abundances from the aforementioned references. The radial oxygen abundance gradient shows a well defined negative slope with a relatively low scatter. We adopt here the following abundance gradients expressions as a function of radius normalised to $\rm R_{25}$,  12+log(O/H) = (8.716$\pm$0.023) - (0.832$\pm$0.044) $\rm R/R_{25}$ for oxygen (see Fig.~\ref{fig:SD_profiles}); and for N/O, log(N/O) = (-0.505$\pm$0.029) - (1.415$\pm$0.075) $\rm R/R_{25}$; the N/O gradient flattens beyond $\rm R/R_{25}$\,=\,0.64 where N/O is better characterised by a constant value of log\,(N/O)\,=\,-1.434$\pm$0.107, reflecting the classical primary to  secondary behaviour of nitrogen.
  
Chemical abundances for NGC\,628 \hii\ regions have been derived in a number of papers. We note the integral field spectroscopy studies performed by \citet{2011MNRAS.415.2439R}, which produced an extended mosaic of the galaxy and, more recently by \citet{2018MNRAS.477.4152R} using SITELLE \footnote{Spectrom\`{e}tre Imageur a Transform\'{e}e de Fourier pour l'Etude en Long et en Large de raies d'Emission} large field fourier transform spectrograph. However in these two works direct derivation of abundances is lacking, as it is also the case of \citet{1998AJ....116..673F} who observed several \hii \ regions beyond NGC\,628 optical radius. The oxygen abundance gradient has been recently derived for the inner $\approx$7 kpc disc of NGC\,628 by \citet{2016MNRAS.455.1218B} though using bright line abundance calibrations. Direct chemical abundances have been obtained by \citet{2013ApJ...775..128B} and \citet{2015ApJ...806...16B} for the largest number of NGC\,628  \hii\ regions with electron temperature measurements to date. The radial abundance gradients of O/H and N/O obtained in the last reference have been adopted in this work. It is surprising the large variance seen in the O/H abundance, which shows scatter along galactic radius. In contrast, the N/O radial gradient appear well defined with a negative slope and low scatter. \citet{2015ApJ...806...16B} suggest various explanations for the scatter such as the selection of the corresponding electron temperature across the ionisation structure of the \hii\ regions, an effect which should be minimised in N/O due to the lower dependence of this abundance ratio on electron temperature. The abundance gradients here adopted are:  12+log(O/H) = (8.835$\pm$0.069) - (0.485$\pm$0.122) $\rm R/R_{25}$ and log(N/O) = (-0.521$\pm$0.035) - (0.849$\pm$0.064) $\rm R/R_{25}$, and beyond the isophotal radius the N/O gradient appears to flatten out, as derived in \citet{2015ApJ...806...16B}. The zero points of M\,101 and NGC\,628 O/H and N/O radial gradients are very similar, to within the errors, though the O/H slope is flatter for NGC\,628 presenting a higher O/H abundance than M\,101 along the disc within the optical radius. In the case of N/O the difference in the gradient slopes seems less enhanced (though still apparent in the same sense). 
The radial profiles of the gaseous abundances, and corresponding uncertainties, adopted for M\,101 and NGC\,628 are shown in Fig.~\ref{fig:SD_profiles}.

\subsubsection{Stars} 

The M\,101 stellar metallicity has been derived by \citet{2013ApJ...769..127L} fitting evolutionary population synthesis \citep{2003MNRAS.344.1000B} models to a set of multiband photometric images (ultraviolet, optical and infrared) together with the fifteen narrowband images observed in the BATC filter system.  
Their metallicity map shows an average radial gradient, giving [Fe/H]$\sim$\,-0.2\,dex in the center and a radial gradient slope -0.011\,dex/kpc. Thus [Fe/H] varies between central -0.2 dex to -0.365 dex at R=15\,kpc for the stellar abundance, with typical uncertainties of 0.2 to 0.3 dex \citep[see Fig. 9 in][]{2013ApJ...769..127L}.
Translating this to oxygen abundance we use the [O/Fe]--[Fe/H] relation, as shown by \citet{2016AN....337..944M}. Taking into account the large uncertainties quoted before for the stellar [Fe/H], and being conservative we have adopted an average correction [O/Fe]=0.15, (for average [Fe/H]$\approx$-0.3), and have applied this correction when deriving [O/H] values from the [Fe/H] measurements, leading to 12+log(O/H) $\approx$ 8.64 at R=0\, kpc and 12+log(O/H) $\approx$ 8.47 at R=15\,kpc. Corresponding gaseous abundances (following the adopted radial gradient, \citet{2016ApJ...830....4C}) are 12+log(O/H) = 8.716 at  R=0, and 12+log(O/H) = 8.32 at R=15\ kpc. The derived stellar metallicity values are smaller than the corresponding gas metallicity in the inner region and slightly higher at R=15\ kpc, but still within the uncertainties of 0.2 to 0.3 dex reported \citep[][]{2013ApJ...769..127L}. The stellar metallicity radial profile, and corresponding uncertainty band, finally adopted for M\,101 is shown in the bottom-left panel of Fig.~\ref{fig:SD_profiles}.

For NGC\,628 we have adopted the stellar metallicity results by \citet{2014MNRAS.437.1534S}. These authors performed a detailed spectroscopic population synthesis \citep[MILES]{2006A&A...457..787S} fitting to the PINGS integral field spectroscopy data of this galaxy from \citet{2011MNRAS.415.2439R}. Spatially integrated spectra of different regions of this galaxy defined via a Voronoi-tessellation binning scheme were fitted. The radial profile for R < 0.5\,${\rm R_{\rm 25}}$ of mass-weighted metallicity for NGC\,628 as derived by \citet{2014MNRAS.437.1534S} has been adopted in this work, and is presented in the bottom-right panel of Fig.~\ref{fig:SD_profiles}.  

\subsection{Gas masses and radial profiles}\label{sec:gasmass}
We use the data from three different surveys: \hi\ observations from 'The \hi\ Nearby Galaxies Survey' \citep{2008AJ....136.2563W} and $^{12}$CO(2-1) from HERACLES \citep{2009ApJ...702..352L,2013AJ....146...19L} to derive \hi\ and $\rm H_{2}$ gas masses, and far-IR observations from KINGFISH to obtain dust masses over the disc of the two galaxies. 

The \hi\ and CO images were convolved to the spatial resolution of 500\,\mi\ image from Herschel using the dedicated kernels from  \citet{2012ApJ...756..138A}, and regridded to the same pixel scale of 14\arcsec. The description of the process is presented in \citet{2013ApJ...777....5S}.  For the \hi\ , we assume the uncertainties to be the larger of either 0.5 M$_{\odot}$/pc$^{2}$ or 10\% of the measured column density, as in \citet{2013ApJ...777....5S}. The CO uncertainties are estimated from the propagation of uncertainties through the different steps involved in creating the integrated CO line map \citep[see][]{2013ApJ...777....5S}. 

\begin{figure*} 
\includegraphics[width=0.3\textwidth]{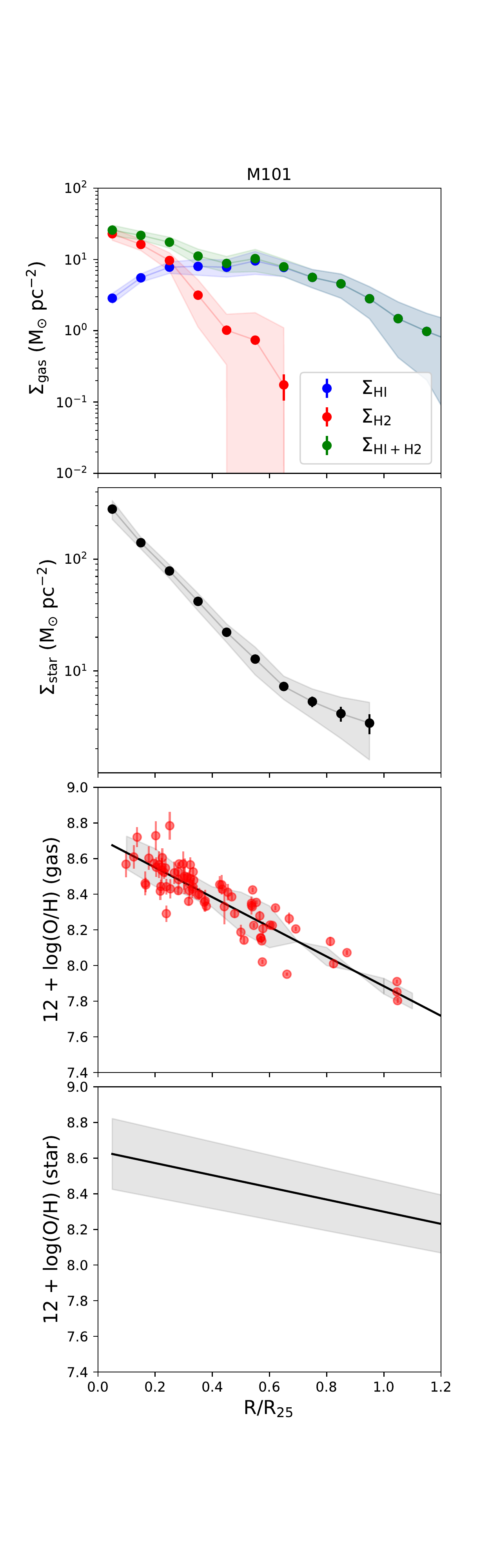}
\includegraphics[width=0.3\textwidth]{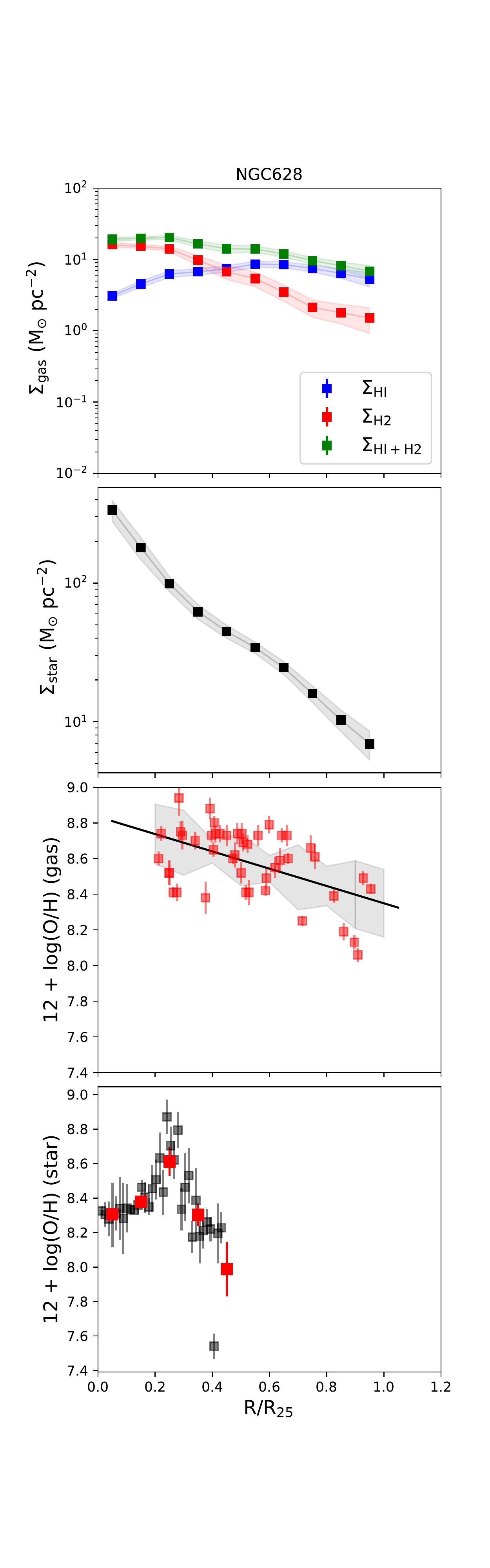}
   \caption{Radial profiles of surface density of \hi\ , $\rm H_{2}$, stellar mass, and of gas and stellar oxygen abundance obtained by binning every 0.1$\rm R_{25}$. \hi\ data have been taken from \citep{2008AJ....136.2563W} and the $^{12}$CO(2-1) observations to derive $\rm\Sigma_{gas}$ are from \citep{2009ApJ...702..352L,2013AJ....146...19L}. Gas oxygen abundances are from \citet{2016ApJ...830....4C} and  \citet{2015ApJ...806...16B} for M101 and NGC\,628, respectively. Stellar metallicities are obtained from \citet{2013ApJ...769..127L}  and \citet{2014MNRAS.437.1534S} for M101 and NGC\,628, respectively. $\rm\Sigma_{gas}$ has been obtained assuming a X$_{\rm CO}$  factor depending on the metallicity at each radial distance. Each data point represents the mean and the uncertainty of the mean for each elliptical ring. The uncertainties in the mean values are generally smaller, when not shown, than the points. The shaded area corresponds to the standard deviation of the values within each ring. In the case of the gas oxygen abundance the shaded area shows the $\pm$1$\sigma$ area above the radial gradient, where $\sigma$ is the standard deviation of the oxygen abundance at each radial bin. See the text for the references of the sources of original data and of previous work on dust content. }
   \label{fig:SD_profiles}
\end{figure*}

In order to derive the amount of neutral gas mass we need to make use of the X$_{\rm CO}$ factor and the line ratio R$_{21}$ =(2--1)/(1--0). For R$_{21}$ we use a value of 0.7, following \citet{2013ApJ...777....5S}. We assume X$_{\rm CO}$ to vary over the disc following the metallicity gradient as X$_{\rm CO}\propto(Z)^{-1}$. The values of X$_{\rm CO}$ derived here agree in general with those obtained by \citet{2013ApJ...777....5S} for both galaxies. A factor of 1.37 is applied to the molecular gas masses to take into account the He contribution. Radial profiles of \hi\ , $\rm H_{2}$, and total gas mass for M\,101 and NGC\,628 are presented in the left and right panels of Fig.~\ref{fig:SD_profiles}, respectively. In the top panels we see that while NGC\,628 presents CO emission up to a radii of $\rm\sim R_{25}$, M\,101 CO emission is restricted to the central $\rm R\leq0.7\,R_{25}$. 
In order to study how the adopted X$_{\rm CO}$ factor can affect the derivation of the molecular gas mass we have applied also a constant X$_{\rm CO}$ factor for the entire discs, scaled to the central metallicity of each galaxy with X$_{\rm CO}\propto(Z)^{-1}$. The radial profiles of the molecular gas mass we have obtained when using a constant X$_{\rm CO}$ across the disc do not change significantly from those presented in the top panel of Fig.~\ref{fig:SD_profiles}. 

\subsection{Stellar mass maps}
We obtain the stellar mass map from the S4G survey,  as derived in \citet{2015ApJS..219....5Q}. These authors apply a methodology that allows to separate the emission from the stars and the dust emission using the 3.6\,\mi\ and 4.5\,\mi\ from \Spi.  
Detailed method followed by these authors to decontaminate hot dust emission in 3.6\,\mi\ is based on the expected colours of the different sources emitting at these bands. The final surface density at 3.6\,\mi\ in units of MJy/sr and decontaminated from dust emission is transformed to stellar surface density mass following Eq.\,6 in \citet{2015ApJS..219....5Q}. In this equation a single mass-to-luminosity ratio of M/L=0.6, using Chabrier IMF, is assumed. Radial stellar profiles for M\,101 and NGC\,628 are presented in the second panel of Fig.~\ref{fig:SD_profiles}. No decontamination of the bulge has been done. Both galaxies belong to the S4G survey and were classified as objects hosting a pseudobulge: some contamination from the bulge can be expected though limited to the inner R$\leq$0.1$\rm R_{25}$ \citep{2015ApJS..219....4S}.  

\section{Results}\label{sec:results}

\subsection{Chemical evolution scheme and effective yield profiles}\label{sec:chemevol}

For the purpose of this work, we have used the scheme of the {\it simple model} (SM) \citep[e.g.][]{1975MNRAS.172...13P,1990MNRAS.246..678E} as a guide to characterize the chemical evolution of our galaxies (although it is well known that it can not provide a complete description of a chemical system). In the SM the system starts with an initial pristine mass of gas and no stars, and metallicity is driven by star formation. The SM assumptions include instantaneous recycling and complete mixing of the ISM, as well as constant stellar yield and IMF. The well-known solution for the SM
\begin{equation}
\rm Z_{gas} = y ln (\mu^{-1})
\end{equation}
relates $\rm Z_{gas}$, the metallicity of the gas expressed in mass fraction of metals, and $\mu$ = $\rm M_{gas}/(M_{gas} + M_{star})$, the gas mass fraction, for the corresponding gas mass, $\rm M_{gas}$, and stellar mass, $\rm M_{star}$, being y the true yield of the  stellar metal production (see below).

A convenient measure of the efficiency of gas enrichment in heavy elements is the "effective yield", $\rm y_{eff}$ \citep{1990MNRAS.246..678E,2007ApJ...658..941D}, defined as
\begin{equation}
\rm y_{eff} = Z_{gas}/ln (\mu^{-1})
\end{equation}
This is a useful indicator that measures how much the metallicity of a galaxy deviates from what would be expected for a system with the same gas mass fraction, but that have evolved in a closed box SM framework, i.e. with no flows of gas (inflow or outflow) allowed. Quoting \citet{1990MNRAS.246..678E} the effective yield $\rm y_{eff}$ of a chemical system is the metal yield that would be derived if that system were assumed to be a closed box SM.  

We have derived the gas mass fraction and stellar mass radial profiles for M\,101 and NGC\,628 and have applied the SM framework to obtain their spatially resolved effective yield, $\rm y_{eff}$. All the relevant quantities were derived spatially resolved for both galaxies. Here $\rm Z_{gas}$ has been obtained directly from the oxygen abundance, 12+log(O/H), and throughout this work we refer to this quantity as the mass fraction of oxygen, calculated as $\rm Z_{gas}$=11.81\,(O/H). The profiles of the gas mass fraction, $\mu$, were obtained using the corresponding radial profiles of both galaxies derived in Section\,\ref{sec:gasmass}. 

From the theoretical side, the metal yield integrated for a single generation of stars, $\rm p$, can be obtained as the IMF mass weighted integral of the stellar yield predicted by theoretical nucleosynthesis models for each star of given mass; in this way the integrated yield of a given element for a single stellar population represents the total amount of newly produced element (e.g. oxygen)  by this population. Since very low mass stars (below 0.8\,M$_{\odot}$) and stellar remnants, in fact, would not contribute with matter to the ISM, we must define the real "return fraction" $\rm R$ of the total stellar mass of a stellar generation which is restored to the ISM  (modulo the assumed IMF). With these definitions the "true" yield can be derived as 
\begin{equation}
$\rm y = p/(1-R)$
\end{equation}
where the quantity $\rm 1-R$ represents the mass that is locked up in stars. Different values of the yield can be achieved depending on the IMFs and stellar nucleosynthesis prescriptions, for the same metallicity range; in the case of oxygen this appears strongly dependent on the set of nucleosynthesis prescriptions and the IMF. In the work of \citet{2016MNRAS.455.4183V} the oxygen yield was computed for a set of three IMFs \citep{1955ApJ...121..161S,2002Sci...295...82K,2003PASP..115..763C} and a set of stellar yields  \citep{2010A&A...522A..32R}; they found the important results that the oxygen stellar yield behaves approximately independent of metallicity and that it has a strong dependence on the assumed IMF. Taking $p$=0.007 from \citet{2016MNRAS.455.4183V} as the lower limit for the integrated yield of oxygen (assuming no rotation \citep{2013ARA&A..51..457N} and one third solar metallicity with $\rm R$=0.3), would lead to a true yield $\rm y$= 0.01 for the metallicity range 0.005 $\le$ Z $\le$ 0.02. Recent work by \citet{2018MNRAS.tmp..302P} included stellar rotation velocity (0, 150 and 300 km/s) finding that the average true yield could be increased by a factor $\sim$\,2 (for a Kroupa IMF, 13 to 120\,M$_{\odot}$). The comprehensive works by \citet{2015MNRAS.451.3693M} and more recently Moll{\'a} et al (2018, in preparation) carried out extensive computations of true yields (including oxygen) for a large set comprising 6 IMFs plus 6 sets of nucleosynthesis prescriptions for massive stars and 4 for low-intermediate mass stars; for each of these possible combinations the yields were computed for 7 metallicities ($\rm Z$ =0.000, 0.0001, 0.0004, 0.004, 0.008, 0.02, 0.05). Examination of their results gives reasonable $\rm y$ values for oxygen typically between $\sim$ 0.002 to $\sim$ 0.01 (i.e. from log$\rm y$$\sim$ -2.7  to log$\rm y$$\sim$ -2), with no dependence on metallicity. 
It is clear that assuming different combinations of stellar nucleosynthesis, IMFs and metallicity lead to a range for the theoretically predicted values of the oxygen true yield. Besides, we should bear in mind that models should assume an effective mass limit for a massive star (e.g. $\ge$ 40 M$_\odot$) to leave a black hole remnant, eventually enclosing a significant part of the nucleosynthesis products.  

From the empirical side, derivations of the oxygen yield can be obtained measuring the  maximum observed (plateau) of the oxygen abundance in spirals discs, which is expected to occur towards the central parts of the most luminous galaxies \citep[e.g.][]{2007MNRAS.376..353P}; these authors find an empirical oxygen yield of $\rm y$ = 0.0035\footnote{Strictly speaking this value would provide a lower limit to the true yield}; they corrected for a fraction (0.08 dex) of the total oxygen assumed to be incorporated into dust grains. Similar empirical values for the oxygen yield have been obtained; and the value $\rm y$  = 0.004, (i.e. log$\rm y$= -2.4), was adopted in the well known work of \citet{2007ApJ...658..941D}. An interesting complementary empirical derivation was performed by \citet{2015MNRAS.450..342K} based on the observed local abundance gradient of OB stars in the Galaxy, which they calibrated with the help of a chemical evolution model. This approach lead the authors to derive an oxygen yield $\rm y$ = 0.00313, assuming $\rm R$ = 0.4, a value in good agreement with the determinations mentioned before, considering the differences in metallicity data source and methods used. In summary, typical empirical estimates of the oxygen yield $\rm y$ cluster between 0.003 to 0.004 (i.e. between log$\rm y$$\sim$ -2.5  to -2.4). Taking into account both, the empirical derivations and the theoretical predictions of the true yield, and their uncertainties as illustrated before, we relax here the expected value of the true yield of oxygen to within the reference range -2.6 $\le$ log $\rm y$ $\le$ -2.2.

\begin{figure} 
\includegraphics[width=0.47\textwidth]{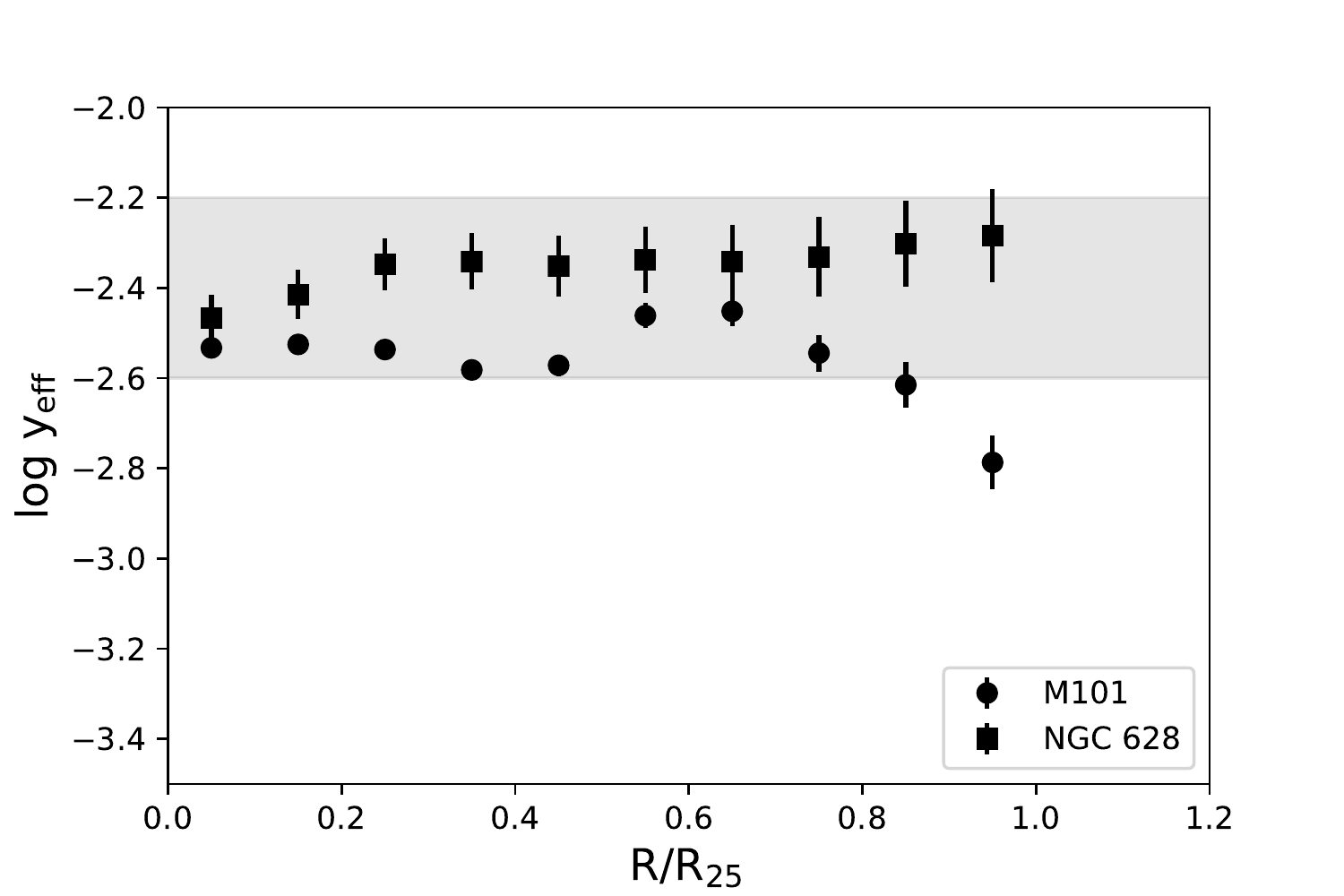}
   \caption{Logarithmic $\rm y_{eff}$ versus galactocentric radius for M\,101 (circles) and NGC\,628 (squares). The grey area corresponds to -2.6 $\le$ log y $\le$ -2.2, the reference band adopted in this work enclosing the expected uncertainties in the derivation of the theoretical true yield. The amount of oxygen depleted onto dust grains has been taken into account assuming a depletion factor of 0.12\,dex \citep[][]{2010ApJ...724..791P}. See the text for details.}
   \label{fig:yeff}
\end{figure}

The radial profiles of the effective yield have been derived for M\,101 and NGC\,628 and they are shown in Fig.~\ref{fig:yeff}. In order to correct the yields for the amount of oxygen depleted onto dust grains, a depletion factor of 0.12\,dex \citep[]{2010ApJ...724..791P} has been adopted; all the effective yield values derived here have been corrected by adding this factor. We can see that  all the effective yield values derived here for both galaxies show log $\rm y_{eff}$ $\le$ -2.2, below the upper face value of the range adopted for the true yield. For all the radii of NGC\, 628 the effective yield derived is within this reference band. This is also the case for a major part of M\,101 ($\rm R/R_{25} \leq$ 0.7) which shows values consistent to within the errors with the reference band, whereas the outer points ($\rm R/R_{25} \geq$ 0.8) of M\,101 present smaller effective yield values clearly departing from the SM picture. This fact illustrates how the 'closed box model' can not provide a good description of the chemical evolution of the entire M\,101 disc for a reasonable expected value of the yield, since the outermost points of its profile strongly suggest the presence of gas flows. Following \citet{2007ApJ...658..941D} the effect of gas inflow is expected to be small, whereas for even moderate gas outflows a substantial  reduction of the effective yield could be observed. Besides that, the N/O radial gradient measured for M\,101 looks very well delineated and a small scatter is shown in the N/O versus O/H plot \citep{2016ApJ...830....4C}, a feature that goes against the presence of massive (external) gas inflow. 

\begin{figure} 
\includegraphics[width=0.47\textwidth]{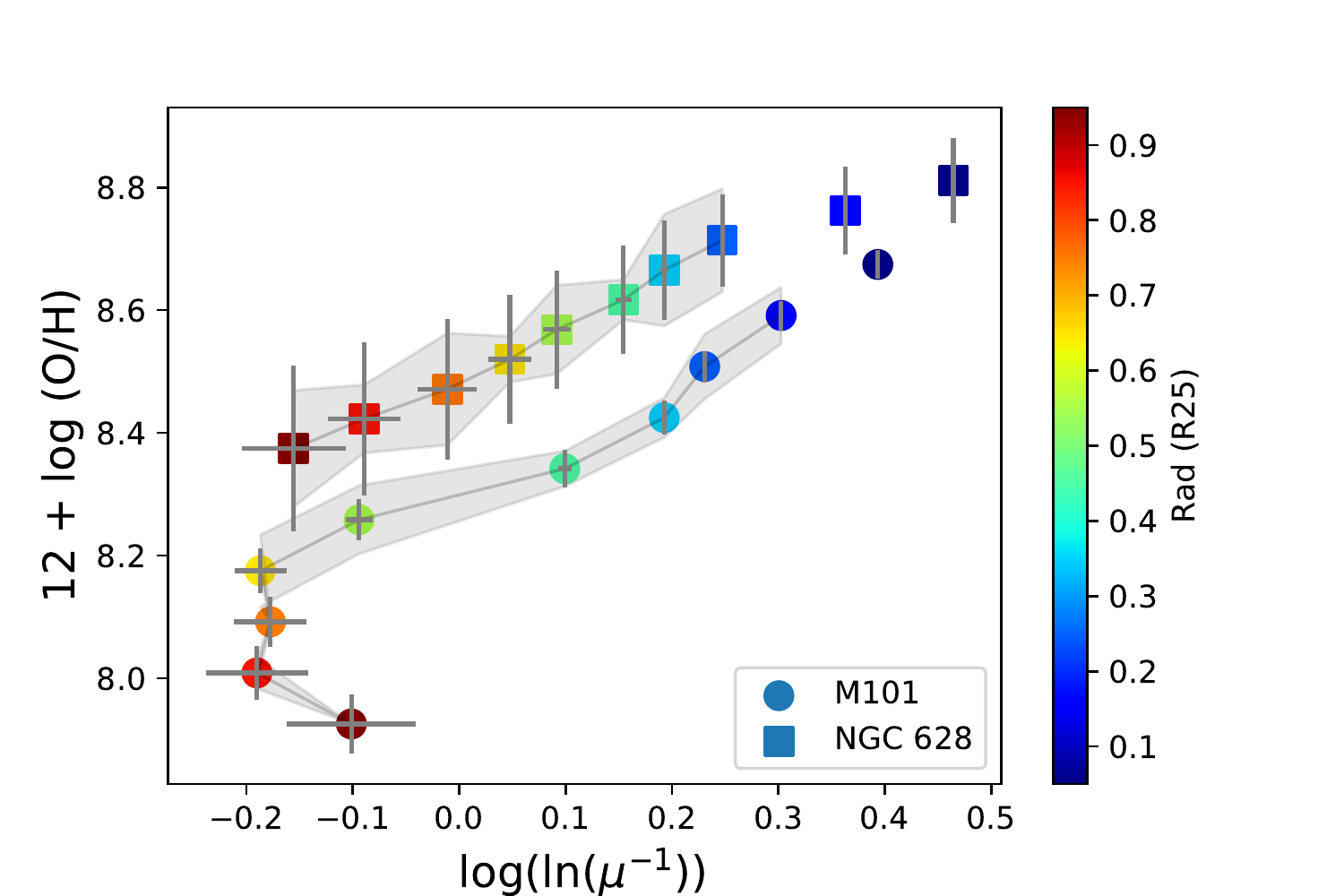}
   \caption{Log-log plot of the oxygen abundance versus $ln(\mu^{-1})$ across M\,101 (circles) and NGC\,628 (squares) for each $0.1\rm R_{25}$ fractional radius.
Points are colour-coded according to the galactocentric radius. A constant effective yield should translate into a constant slope in the plot.}
   \label{fig:logOH_loglnmu}
\end{figure}

In the case of NGC\,628, assuming plausible values of the expected true yield we can accommodate all the observations, thus rendering unnecessary any substantial contribution from gas flows, according to the derived effective yield profile. Nonetheless, we should bear in mind that a moderate unenriched gas infall could not be discarded due to its minimal (detectable) effect in the effective yield \citep[e.g.][]{2007ApJ...658..941D}, as mentioned before. As we can see in Fig.~\ref{fig:logOH_loglnmu}, the oxygen abundance appears well correlated with ln\,$(\mu^{-1})$, the natural log of the inverse gas fraction, for the two galaxies illustrating the locus of the SM of chemical evolution. However for M\,101, for radii R\,$\geq$\,0.7\,$\rm R_{25}$ the observed correlation between oxygen abundance and inverse gas fraction is lost, showing that the 'closed box model' does not seem suitable for the outer parts of this galaxy.  A similar figure for a large sample of low metallicity galaxies has been presented recently \citep{2017MNRAS.471.1743D}. The results obtained we believe are of relevance for the understanding of the behaviour of the metals and dust radial profiles derived for both galaxies, as discussed later in this work.

\subsection{Spatially resolved chemical budget}

The spatially resolved oxygen budget has been computed across the radial profiles of M\,101 and NGC\,628. To do so we first calculate the total amount of oxygen measured along the disc, adding the oxygen content in the stars and in the gas, as well as the correction due to the expected amount of oxygen in dust grains. For each galaxy, the oxygen content in the stars is derived radially as the product of the profile of stellar mass of each galaxy times its corresponding radial profile of stellar oxygen abundance, in mass fraction, as presented in Sect.~\ref{sec:data}. As for the gas, the total oxygen content has been computed for each galaxy as the product -at each radius- of its radial gradient of oxygen abundance of the ionised gas, in mass fraction, times the radial profile of the total gas mass (neutral and molecular; i.e. ionised gas has been neglected). The final oxygen budget has been refined adding a correction accounting for the fraction of oxygen expected to be in dust grains, computed from the expected factor of depletion of oxygen in dust grains in \hii\ regions of $\sim$ 0.12 dex \citep[e.g.][]{2010ApJ...724..791P}.

\begin{figure*} 
\includegraphics[width=0.47\textwidth]{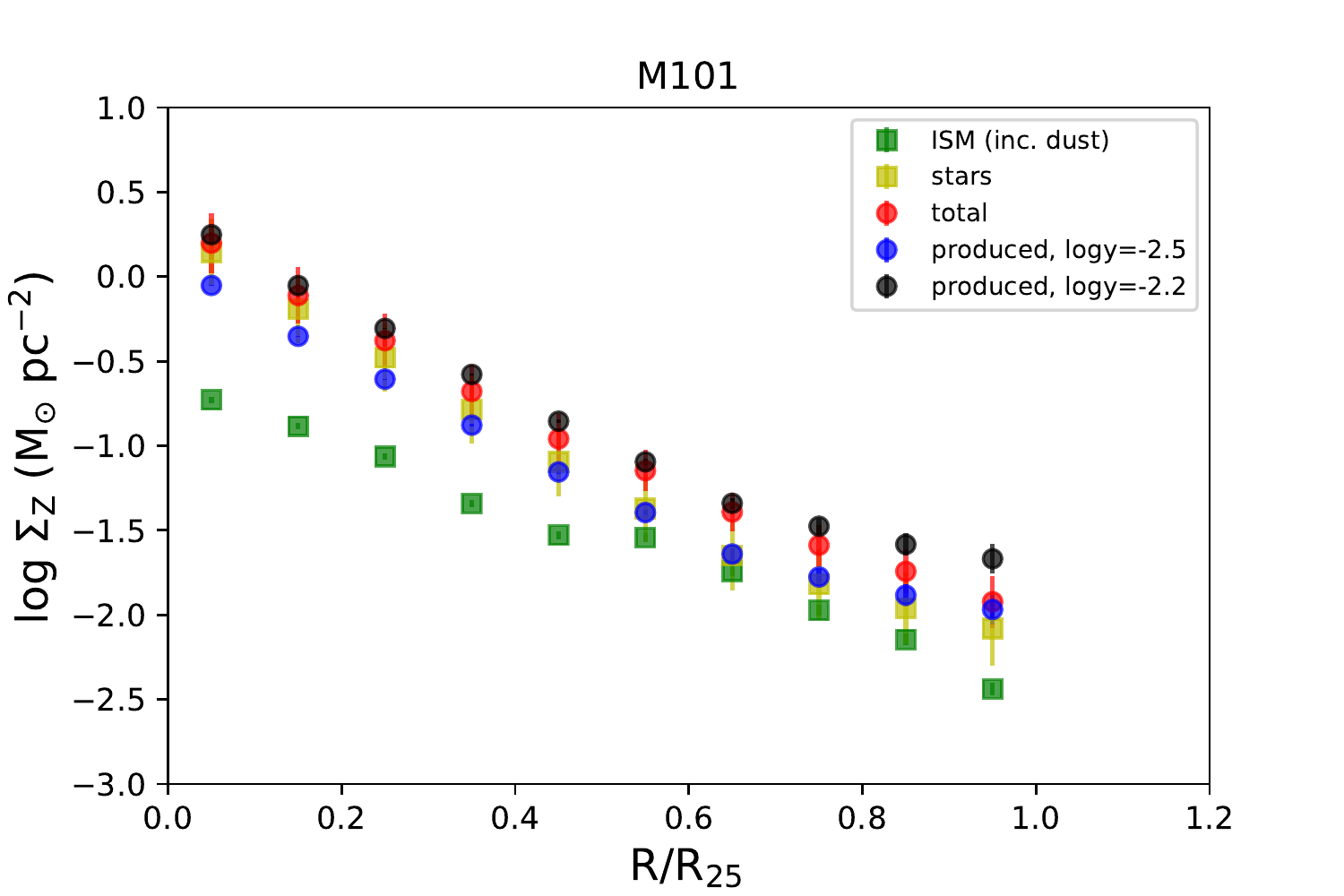}
\includegraphics[width=0.47\textwidth]{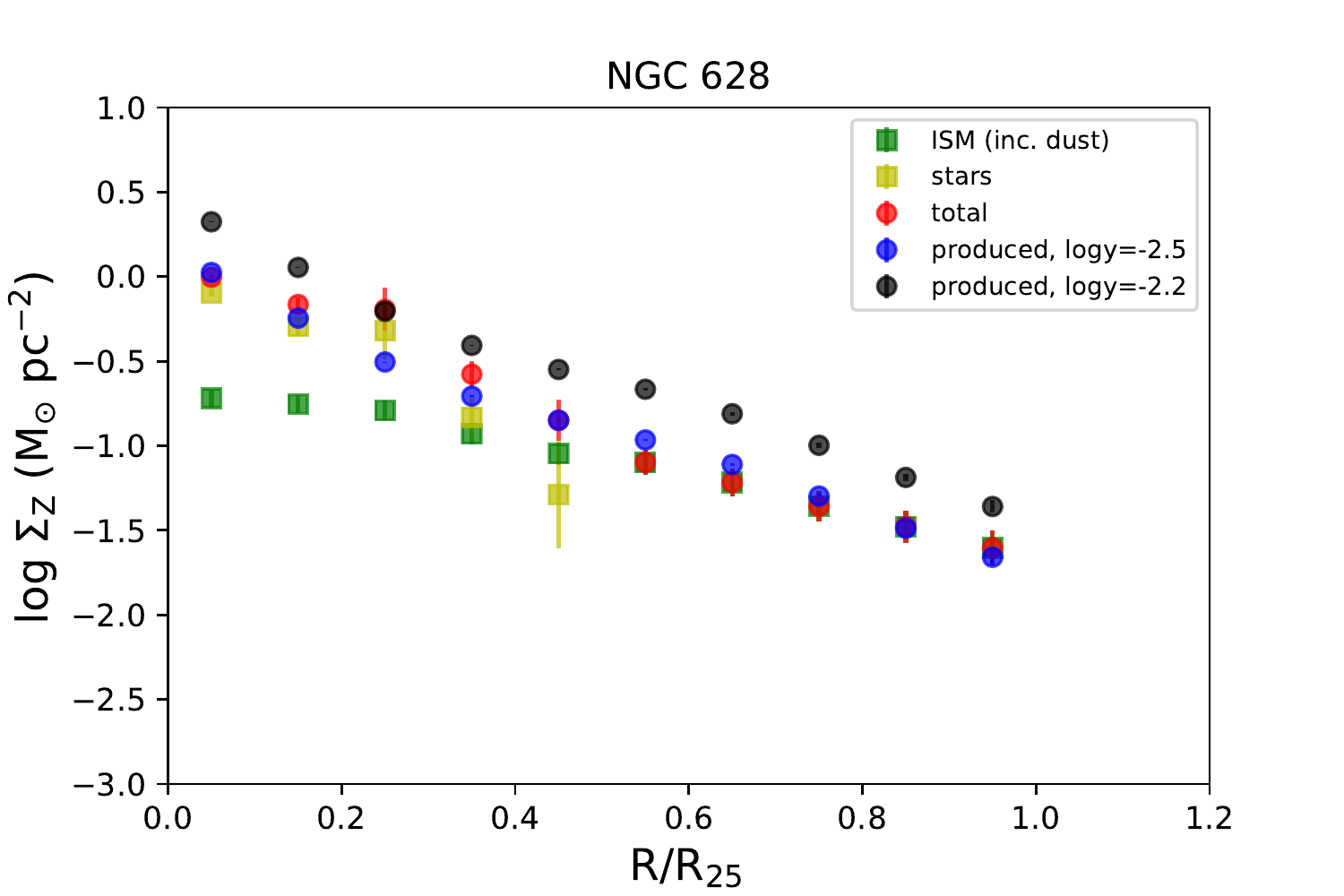}
   \caption{Oxygen budget for M\,101(left) and NGC\,628 (right) as a function of fractional galactic radius. The amount of oxygen in the dust is taken into account assuming a depletion factor of 0.12\,dex \citep[][]{2010ApJ...724..791P}.}
   \label{fig:Obudget}
\end{figure*}

The budget of the "measured" oxygen for each galaxy -'total oxygen' in the plots- can be compared, at each radius, with the expected production of oxygen that has been delivered from stellar nucleosynthesis. This latter bulk stellar production has been calculated at each radius as the product of the (IMF averaged) true yield of oxygen adopted -see previous section-, times the stellar mass radial profile. We show the results in Fig.~\ref{fig:Obudget} for M\,101 (left panel) and for NGC\,628 (right panel). The production of oxygen expected from the stars is presented for two values of the true oxygen yield, log $\rm y$ = -2.2 ($\sim$ solar oxygen) and log $\rm y$ = -2.5 ($\sim$ half solar), for illustrative purposes, encompassing the values of a useful comparison band according to theoretical and empirical predictions.
 
The first result for the figures is that, overall, the radial oxygen budget appears consistent with the observations of both galaxies for the range of values of the true yield adopted. Nonetheless, there are some differences between both galaxies for their oxygen budget comparison. In the case of M\,101, the total oxygen measured is always in between the values of the theoretical estimations of the expected production for the two yields adopted; with the largest contribution to the total measured oxygen coming from the stars, whereas the ISM contribution is nearly an order of magnitude smaller in the inner parts but increases towards the outermost regions of the galaxy. We should bear in mind that we have shown how these outermost regions of M\,101 diverge from the closed box model scenario, and we can see in the plot how the total oxygen for these regions (R\,$\geq$\,0.8\,$\rm R_{25}$) appears closer to the expected production corresponding to the lower true yield adopted (blue points).  

In the case of NGC\,628, the situation appears qualitatively similar, though this galaxy shows a bulk metallicity always higher than that of M\,101 for each fractional radius. However, we can see how the expected oxygen production calculated for the higher true yield adopted (log $\rm y$ = -2.2) is above the measured total oxygen. When a true oxygen yield of log $\rm y$ = -2.5 is assumed the total oxygen measured agrees well with the expected stellar production within the errors for all except one point\footnote{Note that the stellar metallicity contribution is unknown beyond $\rm R/R_{25}$\,$\approx$\,0.5 and could not be added to the total budget}. Of course if a higher value of the oxygen true yield were assumed then a correspondingly larger oxygen production would be predicted and consequently a deficit of oxygen deduced for the galaxy. The radial chemical budget of the inner disc (up to $\sim$ 0.6\,$\rm R_{25}$) of NGC\,628  has been studied in a recent paper by \citet{2016MNRAS.455.1218B} finding that around 50 $\%$ of the metals of the disc were lost, and also that an episode of late time metal enriched gas accretion was present. These results could be reconciled with our findings considering that the oxygen abundances they used are substantially higher and were derived from bright line calibrations (we have used direct oxygen abundances). Also their assumed true yield of oxygen was higher -by 0.2 dex- than the upper boundary (log $\rm y$ = -2.2) of the band for the true yield of oxygen adopted here.   

\subsection{Derivation of dust mass}\label{sec:dustmass}
We derive the dust mass in each pixel of M\,101 and NGC\,628 using the same methodology as described in \citet{2018A&A...613A..43R}. We explain here the main steps of the methodology and refer the reader to \citet{2018A&A...613A..43R} for further technical details. 
We apply the dust model from \citet{1990A&A...237..215D}  and assumed that the spectral shape of the ISRF\footnote{InterStellar Radiation Field} is that of \citet{Mathis:1983p593} to fit the observed SED in each pixel of the galaxy disc. The \citet{1990A&A...237..215D} dust model consists of three dust grain types: polycyclic aromatic hydrocarbons (PAHs), very small grains (VSGs) of carbonaceous material, and big grains (BGs) of astronomical silicates. In order to perform the fit of each individual SED, we first create a library of models with different values of $\rm Y_{i}$ ($\rm Y_{i}=M_{i}/M_{H}$, for i = PAH, VSG, and BG), $\rm G_{0}$ (the scaling factor relative to the solar neighbourhood ISRF given in \citet{Mathis:1983p593}), and $G_{\rm NIR}$ (the scale factor of the NIR 1000\,K black body continuum). Then, we convolve the SEDs of each dust model in the library with the corresponding filter bandpass of our observations to obtain the fluxes in each band for each library model. From the probability density function (PDF) generated in the fit we retrieve the best fit parameter values and the corresponding uncertainties. We take the mean of the PDF as the best parameter value and the 16th-84th percentile range as an estimate of its uncertainty. With the best fit parameters we can derive the dust mass in each pixel of the galaxy as: $\rm M_{dust} = M_{PAH} + M_{VSG} + M_{BG}$. 

\begin{figure*}
\includegraphics[width=0.47\textwidth]{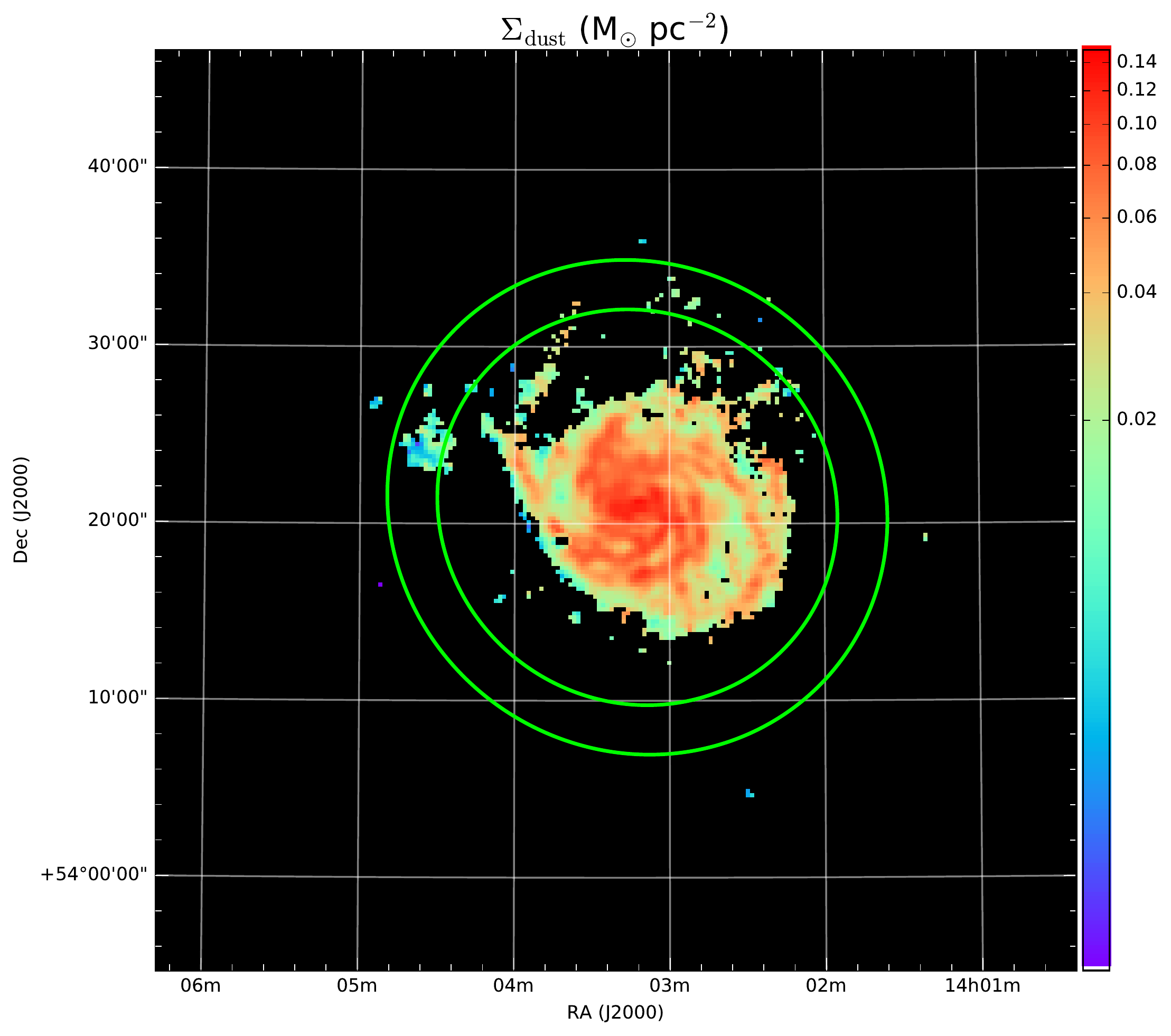}
\includegraphics[width=0.47\textwidth]{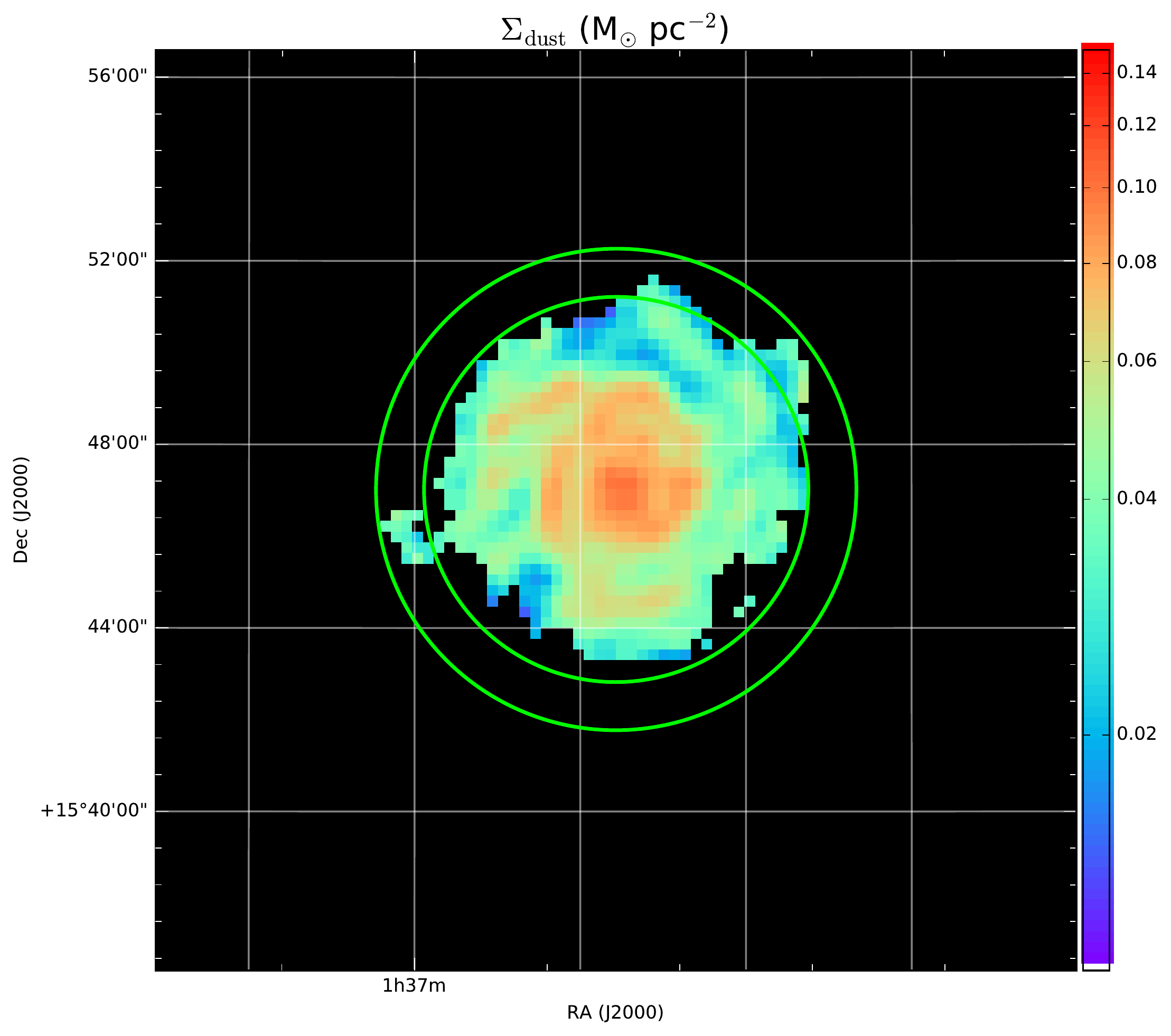}
   \caption{$\rm\Sigma_{dust}$ mass maps for M\,101 (left) and NGC\,628 (right) obtained using the procedure described in \citet{2018A&A...613A..43R}.  In both panels, the ellipses corresponds to $\rm R_{25}$ and 0.8\,$\rm R_{25}$.}
   \label{fig:Mdustmap}
\end{figure*}

\begin{figure} 
\includegraphics[width=0.47\textwidth]{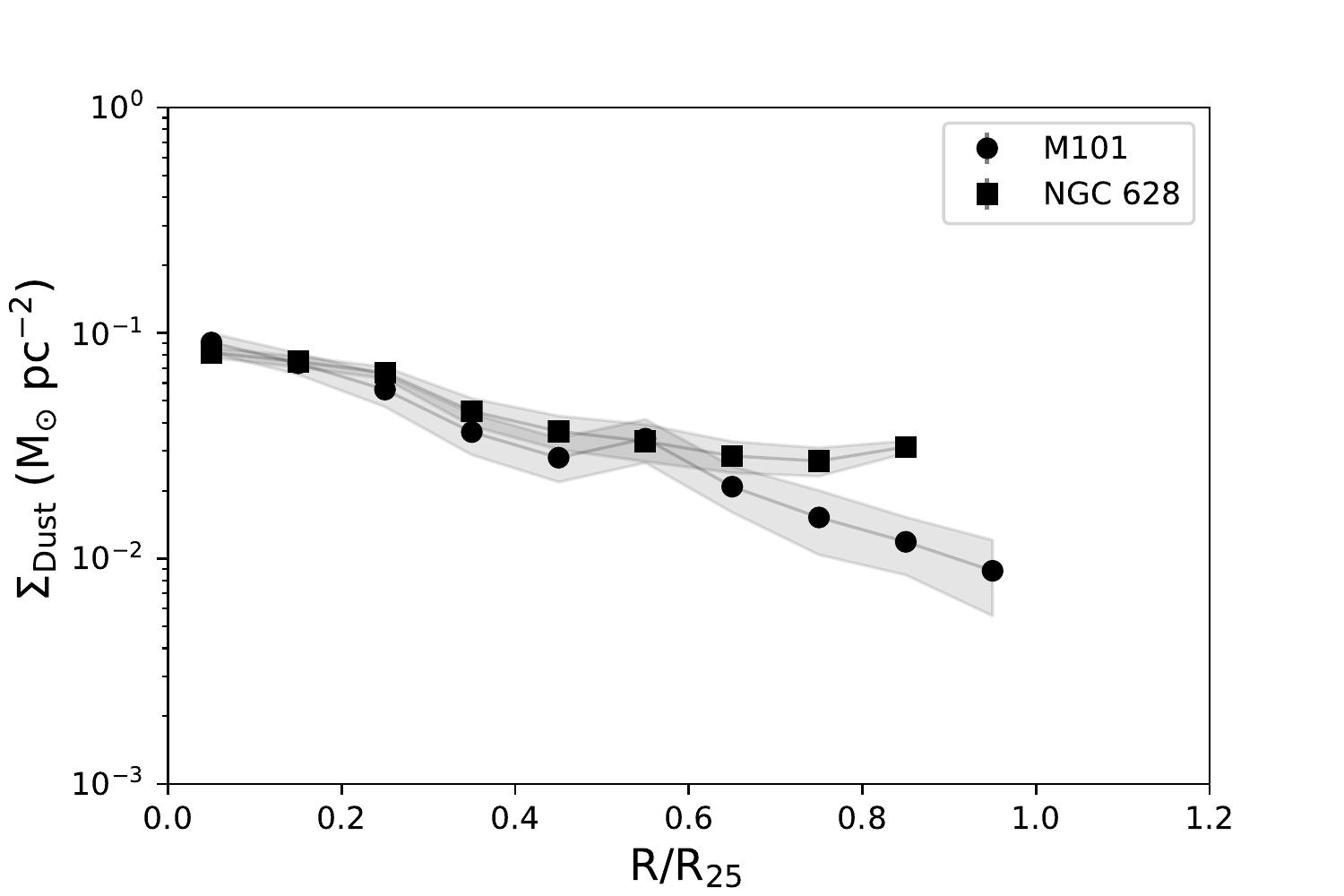}
   \caption{Dust surface density, $\rm\Sigma_{Dust}$, versus radius for M\,101 (circles) and NGC\,628 (squares). As in Fig.~\ref{fig:SD_profiles}, for each data point the mean, and the uncertainty of the mean are presented for each elliptical ring. The uncertainties in the mean values are generally, -not shown-, smaller than the points. The shaded area corresponds to the standard deviation of the values within each ring. }
   \label{fig:SDdust_profile}
\end{figure}

In the fit procedure we have taken only the SEDs with reliable flux measurements in all bands: only pixels with fluxes above 3$\sigma$ in all the bands were taken into account. Finally, we eliminate all the fits with \chidos\,$\geq$\,4.0 and \chidos\,$\geq$\,2.0 for M\,101 and NGC\,628, respectively, in order to obtain reliable dust masses. The relative uncertainties for the dust masses in the final set of pixels are within 10-30\% for both galaxies. The dust masses for NGC\,628 agree with those derived by \citet{2012ApJ...756..138A} when the difference of the extinction coefficients between \citet{1990A&A...237..215D} and \citet{2007ApJ...657..810D} is taken into account. In Fig.~\ref{fig:Mdustmap} we show the dust surface density map for M\,101 (left) and NGC\,628 (right). We see that the map shows internal structures related to the main star-forming complexes within the galaxies and in both objects we are able to derive dust masses up to radii closer to $\rm R_{25}$. n any case, care should be exercised when studying the outer regions of the spiral discs since very low surface brightness, diffuse components of gas and dust could still be present, and they either simply could remain undetected due to their low signal to noise (below detection treshold), or may represent only a tiny contribution in the derivation of radial profiles of the integrated properties.
In Fig.~\ref{fig:SDdust_profile} we show the radial profiles of the dust mass surface density maps for M\,101 and NGC\,628. Dust mass radial profile for NGC\,628 has been previously derived by \citet{2009ApJ...701.1965M} using only \Spi\ bands (e.g. up to 160\,\mi). They find higher dust masses in general and a steeper radial profile than what is found here. Our dust mass estimations are more reliable as they are obtained with data that cover the whole IR SED peak.  

\subsection{Derivation of the gas to dust ratio}
\begin{figure*} 
\includegraphics[width=0.47\textwidth]{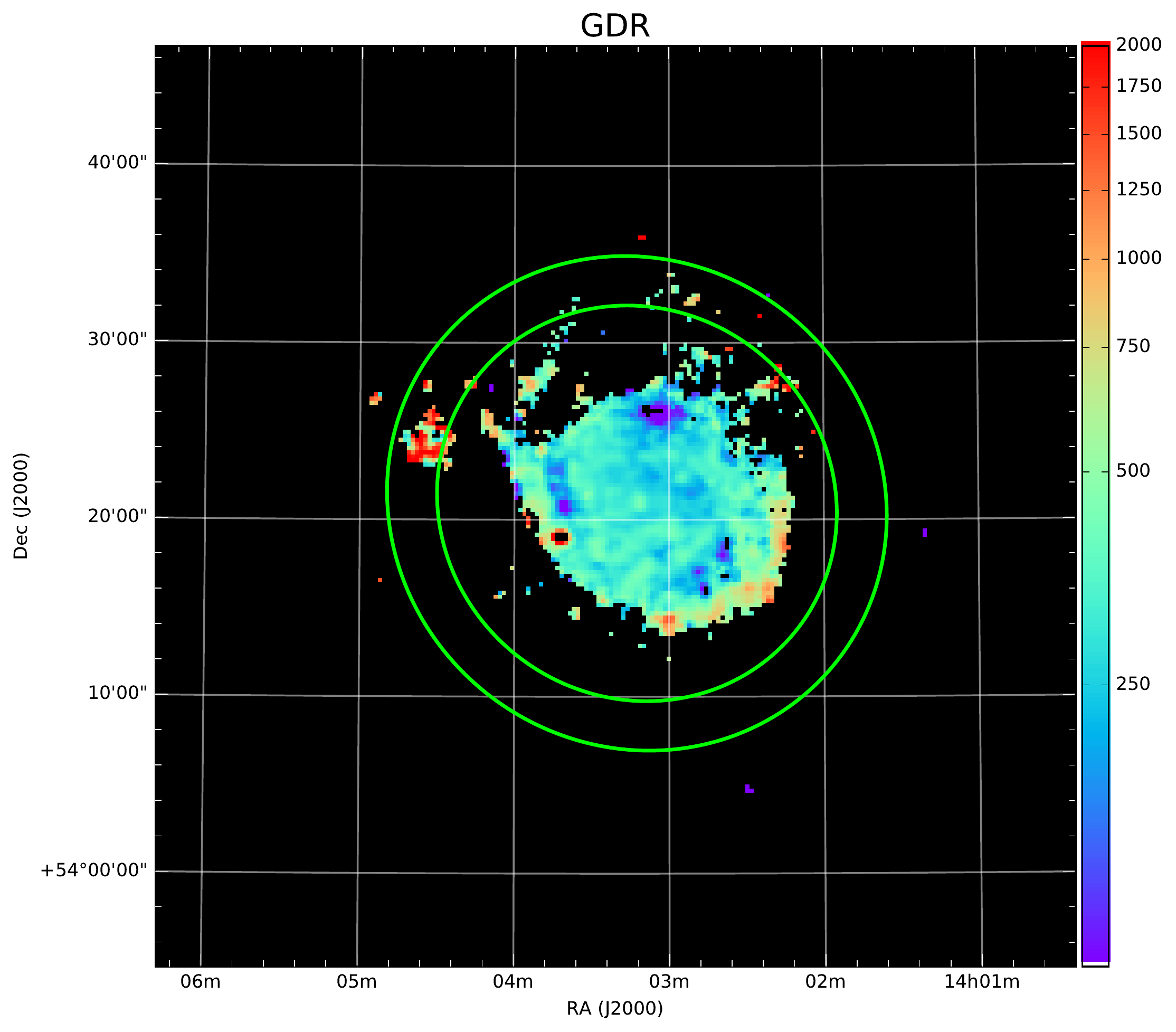}
\includegraphics[width=0.47\textwidth]{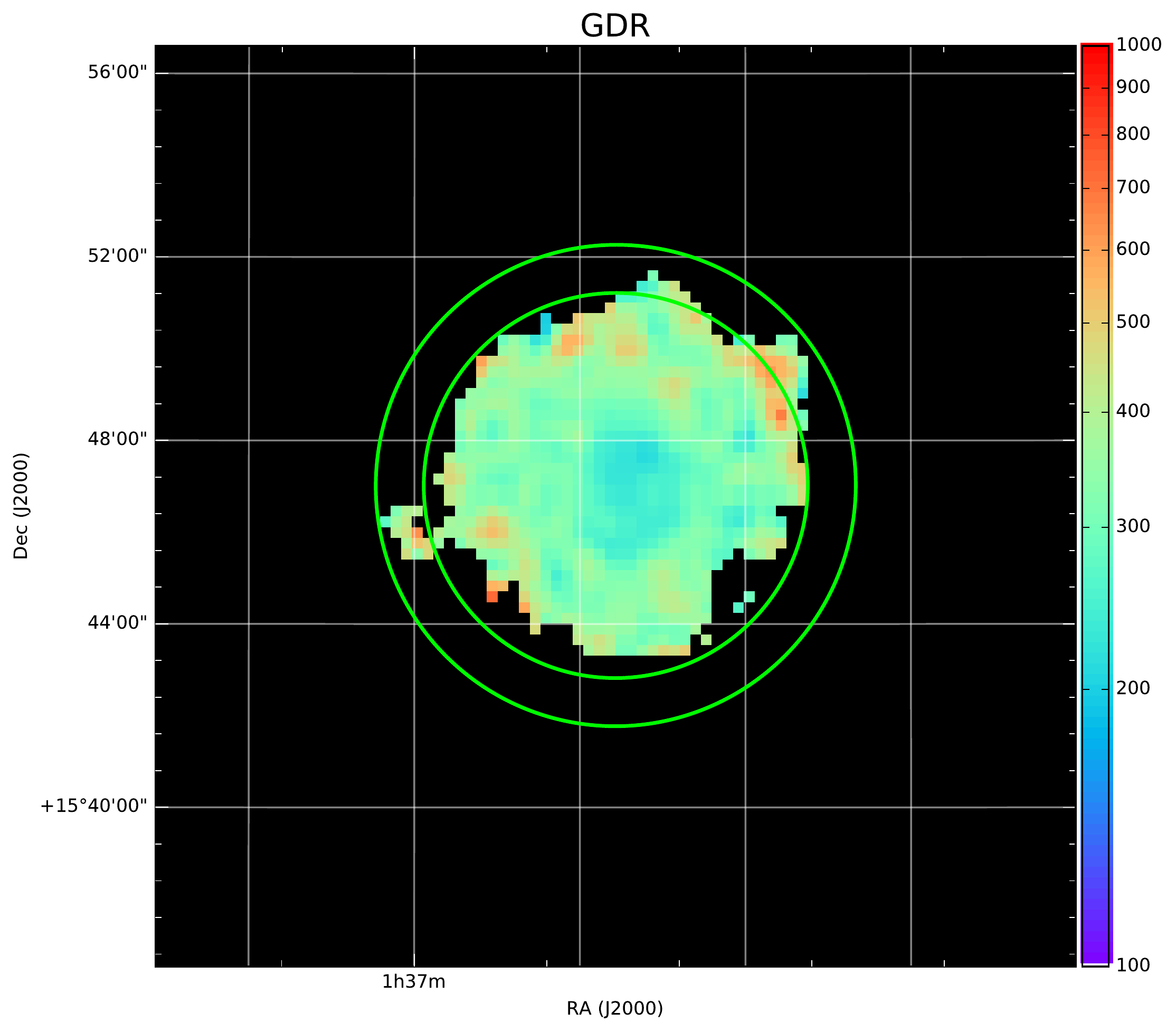}
   \caption{GDR maps of M\,101 (left) and NGC\,628 (right). In both panels, the ellipses corresponds to $\rm R_{25}$ and 0.8\,$\rm R_{25}$.}
   \label{fig:GDRmaps}
\end{figure*}

\begin{figure} 
\includegraphics[width=0.47\textwidth]{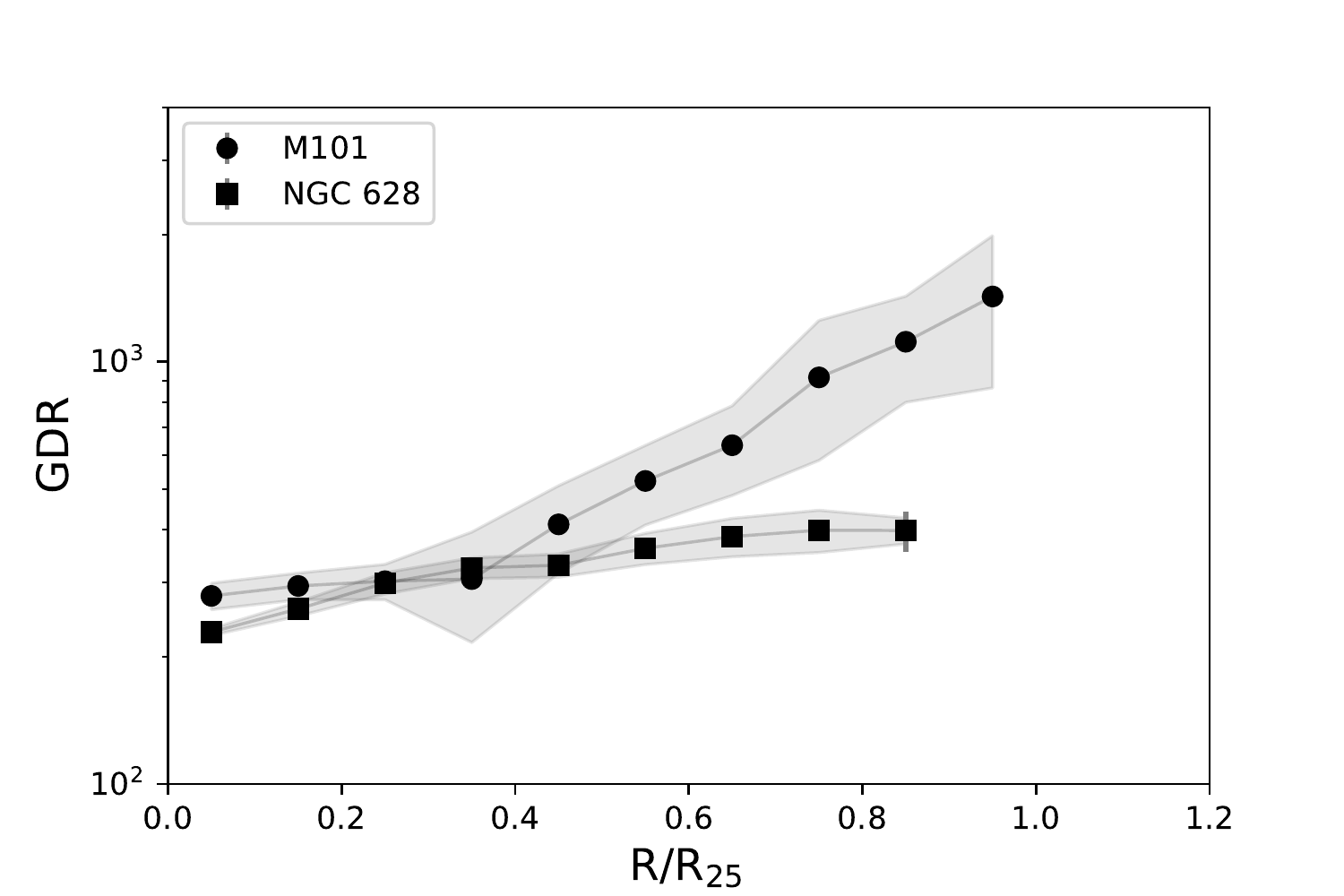}
   \caption{GDR radial profiles for M\,101 and NGC\,628 obtained with the dust masses derived in this paper. Each data point represents the mean and the uncertainty of the mean for each elliptical ring. The uncertainties in the mean values are generally smaller, when not shown, than the points. The shaded area corresponds to the standard deviation of the values within each ring.}
   \label{fig:GDR_profile}
\end{figure}

The gas-to-dust mass ratio (GDR) map for each galaxy is obtained from the total gas mass and the dust mass maps derived in Sections\,\ref{sec:gasmass} and \ref{sec:dustmass}.  In Fig.~\ref{fig:GDRmaps} we show the GDR maps for M\,101 (left) and NGC\,628 (right).  While the GDR of NGC\,628 seems to be relatively constant across the disc of the galaxy, the GDR of M\,101 shows a significant increase towards the outer parts of the galaxy. A radial profile of the GDR map for each galaxy presented in Fig.~\ref{fig:GDR_profile} shows the different behaviour with galactocentric radii for both galaxies. The GDR of NGC\,628 increases very smoothly with galactocentric radius, while the GDR of M\,101 shows a significant increase towards the outer parts of the galaxy. The values of the GDR for both galaxies agree with the values reported by \citet{2013ApJ...777....5S} after taking into account the different extinction coefficients to derive the dust masses. These authors also found significant high values of GDR in the outer parts of M\,101, while for NGC\,628 the GDR map they present is more smooth. The GDR radial profile of M101 presented here also agrees with the one presented in \citet{2018ApJ...865..117C}. 

\section{Discussion}\label{sec:discussion}
\subsection{GDR relation with metallicity and molecular content}
In order to study the chemical evolution and the dust content of M\,101 and NGC\,628 we have derived the radial profiles of the total gas mass and stellar mass of the galaxies extracted from the 2D data, as well as the maps of the dust mass and of the GDR (Figs.~\ref{fig:Mdustmap} and~\ref{fig:GDRmaps}, respectively). These maps show a clear radial trend and also considerable spatial structure going from the inner parts to the outer disc. In both cases the outermost ring beyond 0.8\,$\rm R_{25}$ appears less populated with emission regions, though some big star-forming complexes (like NGC\,5471 in M\,101) are still present as well as more sparse emission. The results obtained for both galaxies have been derived in a consistent way so as to be compared independent of the precise assumptions adopted (e.g. derivation of the molecular gas and of dust components). 

In Fig.~\ref{fig:GDR_profile} we can see how the two galaxies show quite different radial profiles of the GDR. Whereas for NGC\,628 the profile appears nearly flat moving between GDR\,$\sim$\,200 to 400 for the whole range of galactic radius, the case of M\,101 presents an apparently broken radial profile shape starting from GDR values similar to the  ones derived for the inner disc of NGC\,628, however beyond 0.5\,$\rm R_{25}$ the GDR profile becomes very steep reaching well above GDR $\approx$1000 in the outermost disc region. 

The chemical evolution of both galaxies would be consistent with the SM as judged from their radial profiles of the effective yield of oxygen if we assume a lower band for the true yield between solar to half solar  (-2.5\,$\leq$\,log\,y\,$\leq$\,-2.2), and allowing for a 0.12\,dex correction for oxygen depletion onto dust grains. However, while this situation appears to hold for the complete disc up to the optical radius for NGC\,628, for M\,101 applies till $\rm R/R_{25}$\,$\approx$\,0.7. Beyond this radius the derived profile of the oxygen effective yield of M\,101 suggests that a moderate gas outflow could be invoked there. Some small inflow component could not be discarded but its effects should be hardly noticeable, especially given the higher gas fraction already present in its outer disc \citep[e.g.][]{2007ApJ...658..941D}. Massive inflow does seem less likely judging from the well defined N/O versus O/H relation observed in this galaxy\footnote{N/O is not expected to be much affected by massive unenriched inflow, but O/H would be.}. These results qualitatively agree with previous findings in the literature \citep[][]{2015MNRAS.450..342K,2015AJ....150..192Z}, though in the case of NGC\,628 an important loss of metals has been proposed especially for the inner regions \citep[][]{2016MNRAS.455.1218B}. Nonetheless, we should bear in mind that these previous works have used oxygen abundances derived with bright line abundance calibrations that can suffer from well-known important uncertainties (see Sect.~\ref{sec:data}). 

\begin{figure*} 
\includegraphics[width=0.47\textwidth]{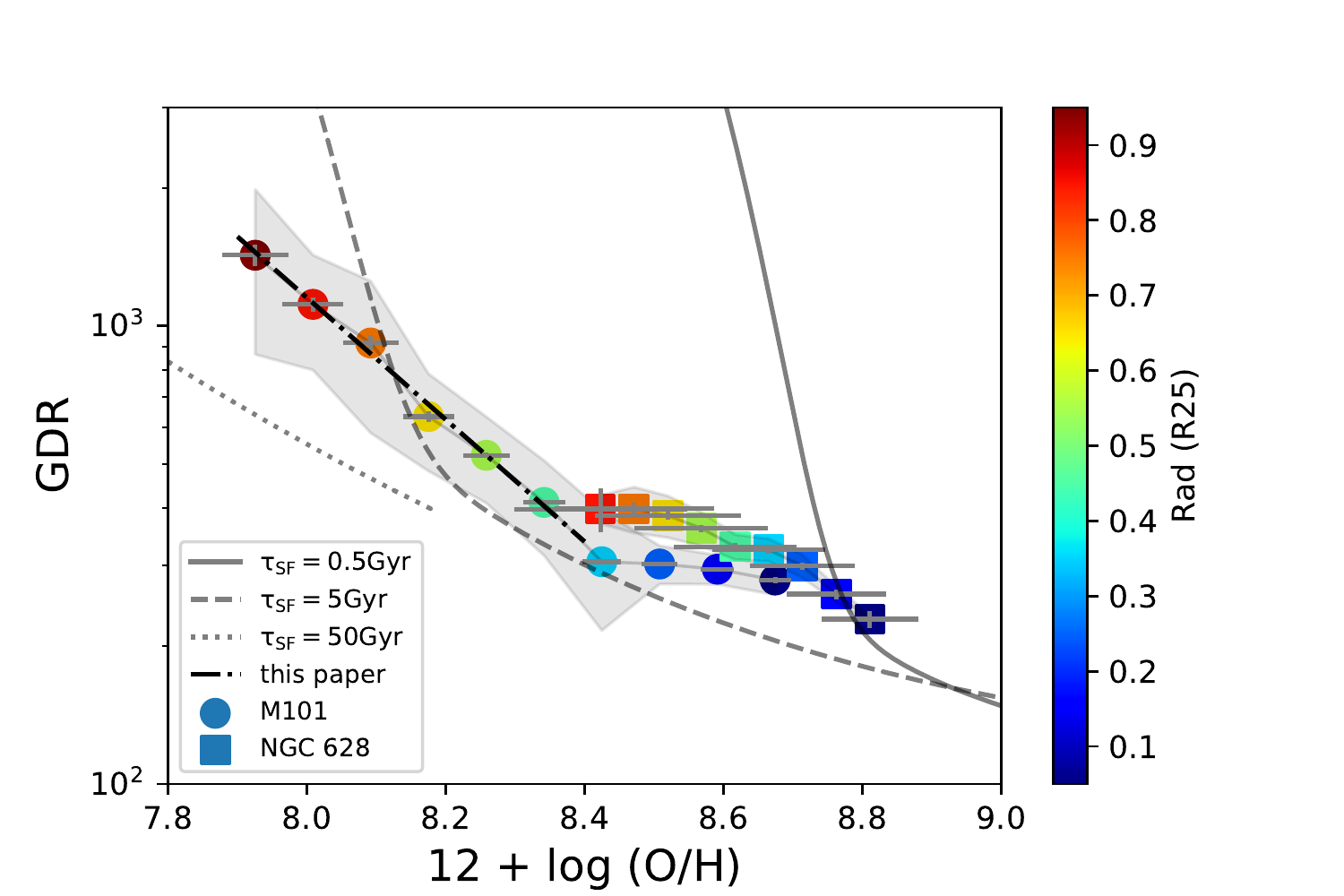}
\includegraphics[width=0.47\textwidth]{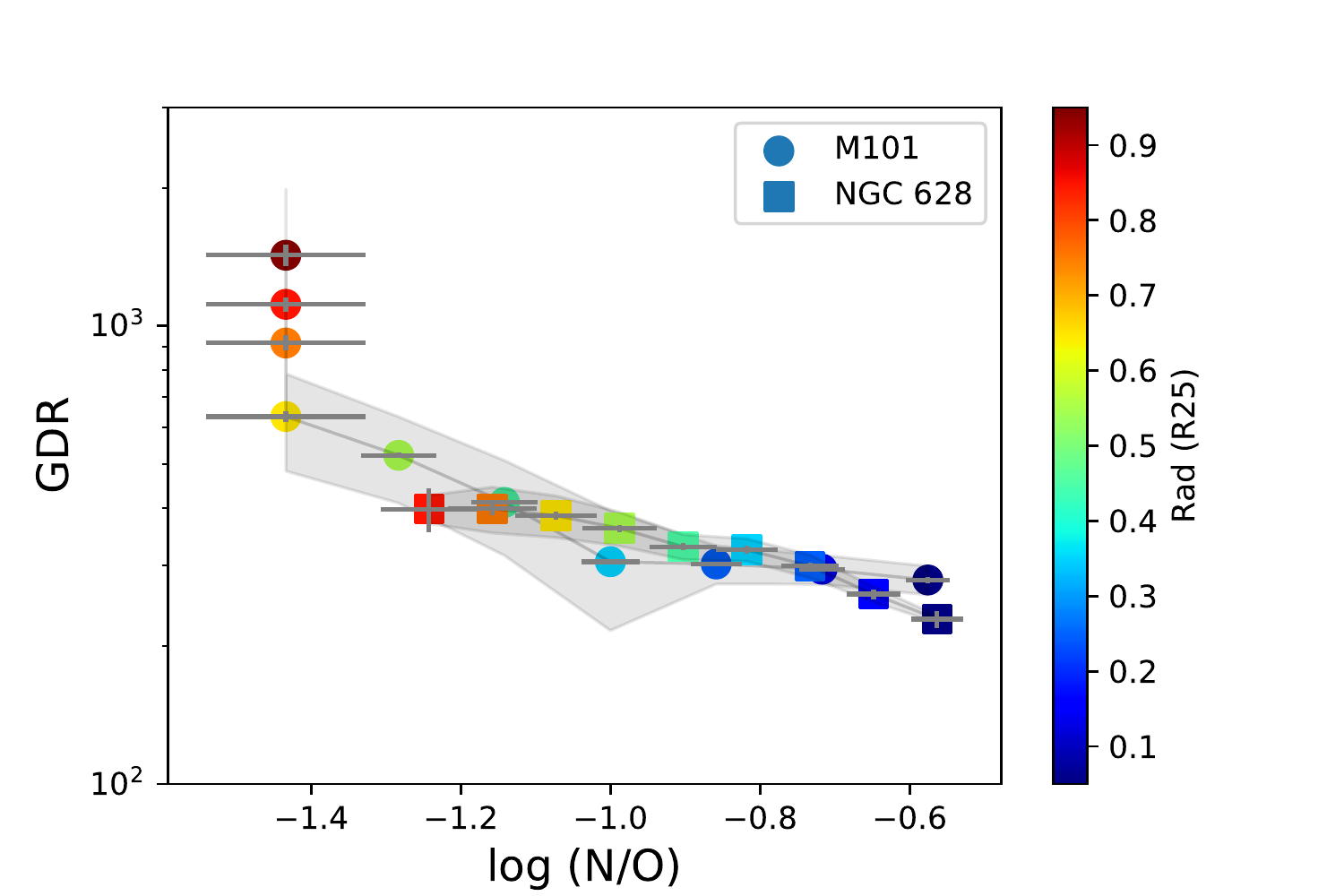}
   \caption{Left: GDR versus the oxygen abundance for M\,101 (circles) and NGC\,628 (squares). The continuos, dashed and dotted lines are the models from \citet{2013EP&S...65..213A} for star formation time scales of 0.5, 5, and 50\,Gyr, respectively. The dot-dashed line is the linear fit to the low metallicity regime shown in Eq.\,3. Right: GDR versus the N/O ratio for M\,101 (circles) and  NGC\,628  (squares). Color code is the galactocentric distance normalised to $\rm R_{25}$.}
   \label{fig:GDR_metal}
\end{figure*}

The radial GDR profile and the spatially resolved effective yield are compared across both galaxies, sampling one dex range in O/H. Here we empirically present the relationship between the gas to dust mass ratio and chemical abundance obtained across both galaxies. In the left panel of Fig.~\ref{fig:GDR_metal} we show the behaviour of the GDR versus the oxygen abundance, 12+log(O/H), for M\,101 (circles) and NGC\,628 (squares).  A clear two slopes behaviour is seen with a break at 12+log(O/H)\,$\approx$\,8.4; below this value of oxygen a steep negative slope  ($\Delta$ logGDR/$\Delta$ log(O/H)$\approx$ -1.3) is present, mostly defined by the external (metal poor) points of the disc of M\,101. For higher metallicities the slope is still negative but much shallower, and it seems similar for both galaxies to within the errors. 

The observational trend shown here appears consistent with that one found by \citet{2014A&A...563A..31R} for a sample of galaxies. These authors present a broken power law to describe the observational behaviour of the GDR versus a single value of 12+log(O/H) for a large sample of galaxies. Their break at 8.10$\pm$0.43 appears consistent, within the errors, with the break at 12+log(O/H)\,$\approx$\,8.4 found in this work. 
 \citet{2014A&A...563A..31R} found a break when a constant Galactic X$_{\rm CO}$  factor was used, as well as when a X$_{\rm CO}$ factor scaled with the metallicity of the galaxy was applied to each object. Here we have used a variable X$_{\rm CO}$ factor that scales with the metallicity (measured) at each galactocentric radius across the galaxy. We have also derived the GDR using a constant X$_{\rm CO}$ factor for the whole disc, scaled with the central metallicity, of each galaxy (see Sect.~\ref{sec:gasmass}). In this case, the radial variations of GDR obtained for both galaxies show the same trend as in Fig.~\ref{fig:GDR_profile}.   
Our results are in good agreement with the findings of the recent detailed work across the M\,33 disc done by \citet{2018A&A...613A..43R}. In the left panel of Fig.~\ref{fig:GDR_metal} we show the theoretical predictions of \citep{2013EP&S...65..213A} which define a critical metallicity highlighting this change in slope. We can see that our observations agree with the chemical evolution models of \citet{2013EP&S...65..213A} for star formation time scales of 0.5-5\,Gyr. 

In the right panel of Fig.~\ref{fig:GDR_metal} the derived GDRs for M\,101 (circles) and  NGC\,628  (squares) are shown against the corresponding log(N/O) ratios across their discs. A strong correlation is present for both galaxies within the metallicity range typically associated to secondary nitrogen production, i.e. for log(N/O) $\geq$ -1.4 (above this N/O value, the break in GDR vs. oxygen abundance -as seen in the left plot- corresponding to log(N/O) $\sim$ -1.1), whereas no correlation is shown below this N/O value, sharing all the points the same average N/O, characteristic of primary nitrogen production at low metallicity.

\begin{figure} 
\includegraphics[width=0.47\textwidth]{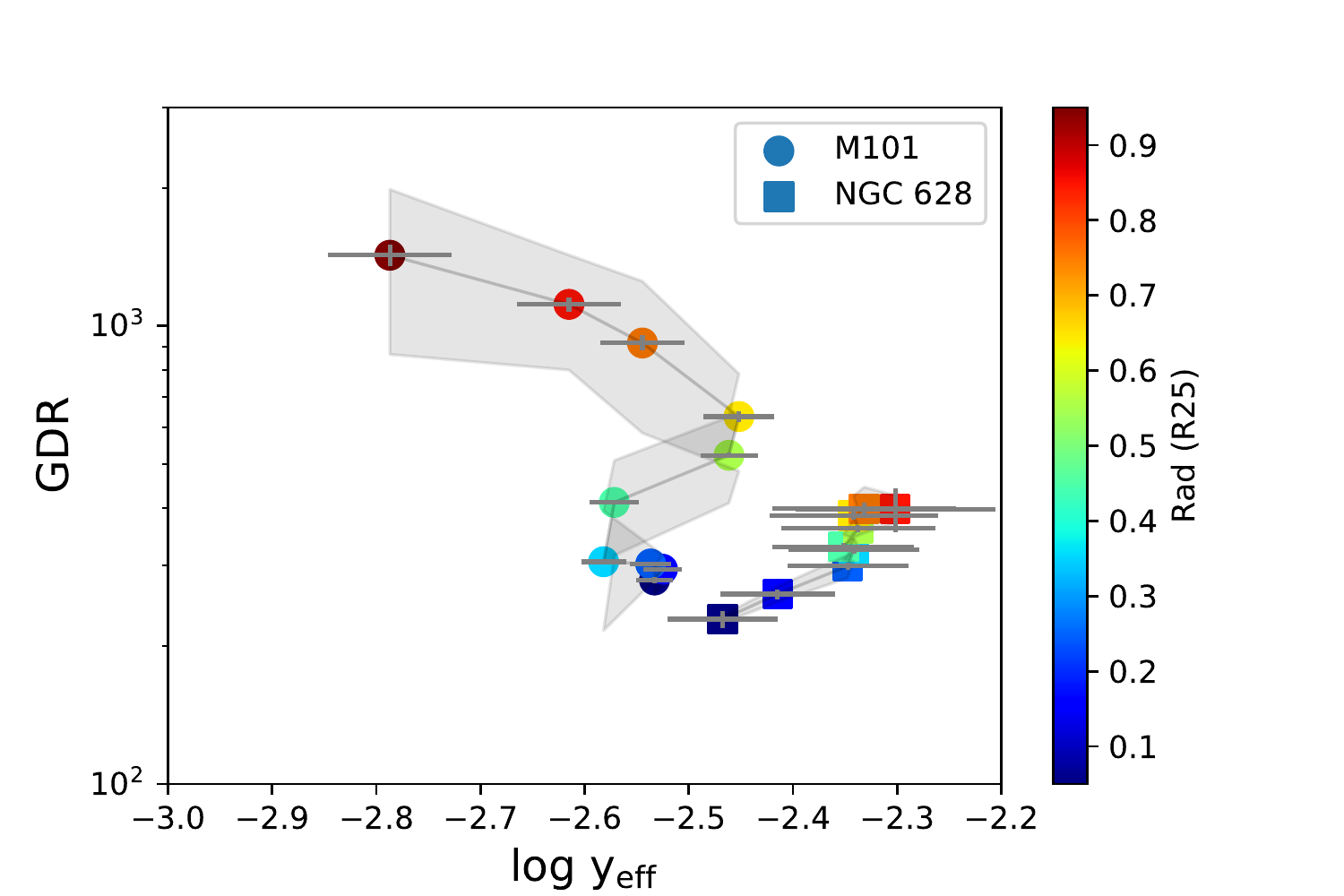}
   \caption{GDR versus the logarithmic of the effective yield for M\,101 (circles) and NGC\,628 (squares).}
   \label{fig:GDR_yeff}
\end{figure}

The effects of the different chemical evolution and the spatial profiles of the oxygen effective yield derived for both galaxies could be translated onto an {\it effective dust yield} for stellar dust production and onto the dust growth component of the rich ISM, especially in the inner galaxy. In this scenario, the largest GDR values obtained in this work have been derived for the outer disc of M\,101, where the largest deviations from the SM yield are measured. In Fig.~\ref{fig:GDR_yeff} we show the GDR derived across both galaxies versus the corresponding values of the effective yield of oxygen.
We can see how the lower values of the GDR derived for M\,101 and the GDR for the entire disc of NGC\,628, all between 200\,<\,GDR\,<\,500, correspond to those regions of the galaxies presenting an effective yield broadly consistent with the SM of chemical evolution. Conversely, it is for the lowest values of the effective yield that we see the highest GDR values derived for the outermost disc of M\,101. 

In the framework of the \citet{2013EP&S...65..213A} models, a two slopes behaviour is predicted for the GDR against 12+log(O/H); these models suggest the existence of a critical metallicity, when stardust production equals grain growth, for which a change in the slope of GDR against metallicity should be observed. Asano's models predict the slope to be steeper in the critical metallicity regime (0.05$\leq$$\rm Z/Z_{\odot}$$\leq$0.3) defined in \citet{2017arXiv171107434G} as observed here. Hints for the existence of an overall behaviour of GDR versus metallicity have been shown in previous studies using diverse samples of galaxies \citep[e.g.][]{2013ApJ...777....5S,2014A&A...563A..31R}, with a somewhat large scatter probably mostly associated to the different sample definition and diverse origin of the individual objects \citep[see e.g.][]{2017arXiv171107434G} included; these two branches have been observed also for  M\,33 by \citet{2018A&A...613A..43R}. More recent models have been performed by \citet{2014A&A...562A..76Z} reproducing this relation of GDR versus metallicity and adding the effects of episodic star formation histories on the ISM dust growth.

\begin{figure} 
\includegraphics[width=0.5\textwidth]{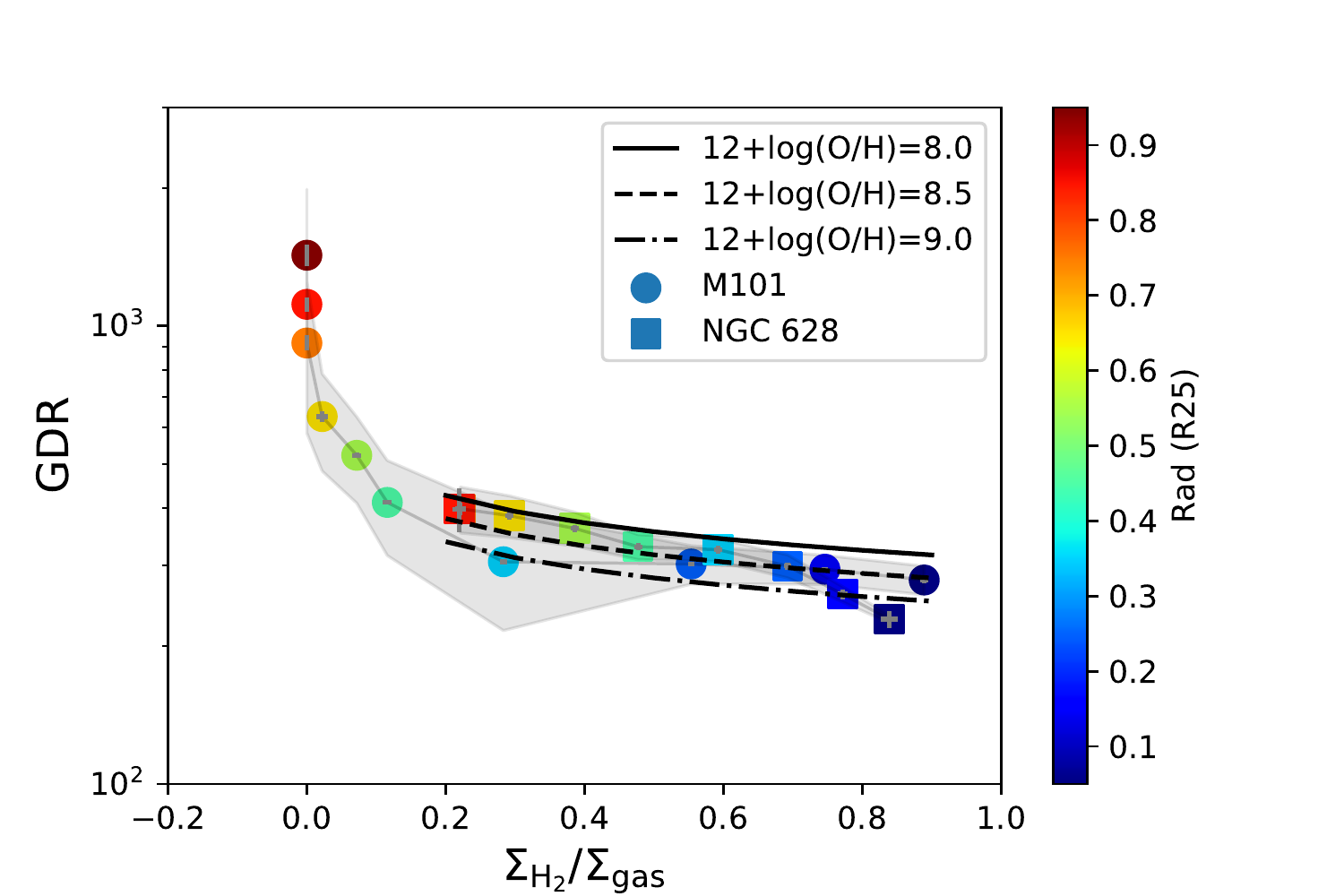}
   \caption{The behaviour of the gas to dust ratio GDR against the cold ISM fraction as mapped by $\rm\Sigma_{H_{2}}$/$\rm\Sigma_{gas}$. Points are colour coded for M\,101 (circles), and NGC\,628 (squares).}
   \label{fig:GDR_fracmolgas}
\end{figure}

The growth of dust grains in the ISM is needed to balance the overall dust budget, adding to the stardust production and compensating by the dust sink of grain destruction which is believed to be associated to SN blast waves (likely also to a hard radiation field), also allowing for the possible net dust loss linked to gas outflows. Within the above mentioned theoretical framework for the evolution of dust content it is predicted that the accretion time for grain growth is expected to be proportional to the inverse of the metallicity and to the inverse of the fraction of cold (molecular) component of the ISM. We have defined a proxy to trace this fraction of cold ISM component using the molecular to total gas mass ratio, $\rm\Sigma_{H_{2}}$/$\rm\Sigma_{gas}$, across the galaxies. In Fig.~\ref{fig:GDR_fracmolgas} we present the behaviour of the derived GDR against the cold ISM fraction mapped by  $\rm\Sigma_{H_{2}}$/$\rm\Sigma_{gas}$ across both galaxies. The colour coded points help us follow their radial positions across the discs. The GDR shows a close relation with $\rm\Sigma_{H_{2}}$/$\rm\Sigma_{gas}$, in this case, it seems to be complementary to what is shown in Fig.~\ref{fig:GDR_metal}; i.e. it is now the lower GDR values (GDR\,$\lesssim$\,400) that appear very well sampled and extended along the  $\rm\Sigma_{H_{2}}$/$\rm\Sigma_{gas}$ axis.  Both galaxies are seen to share a strong correlation with a shallow negative slope above $\rm\Sigma_{H_{2}}$/$\rm\Sigma_{gas}$\,$\approx$\,0.1. For the regions with a very low molecular fraction, mainly populating the outer disc of M\,101, the GDR vs. $\rm\Sigma_{H_{2}}$/$\rm\Sigma_{gas}$ slope is nearly vertical in the plot, witnessing a strong non dependence of the high GDR values of these regions on the ISM grain growth.

The observed trend of GDR versus $\rm\Sigma_{H_{2}}$/$\rm\Sigma_{gas}$ has been linked to grain growth, which should take place in regions of the ISM of intermediate to high metallicity  \citep[][]{2017arXiv171107434G}. This fact can be interpreted as a straight (causal) effect, but also a complementary scenario could be invoked in which a significant amount of hydrogen molecules can form on the surfaces of the dust grains; and thus, an enhanced $\rm\Sigma_{H_{2}}$/$\rm\Sigma_{gas}$ relative content should be found associated to those regions with more dust grains and high surface dust density.
 
We have empirically unveiled here a twofold behaviour for the GDR: the gas-to-dust ratio is seen to correlate with metallicity (12+log(O/H)), and also with the molecular fraction of the total gas content. A clear change in the observed slopes is seen for: a) 12+log(O/H)\,$\sim$\,8.3 to 8.4, the critical metallicity predicted in Asano's models; this value appears to be similar for the two discs studied here, though for NGC\,628 the points do not enter into the steep slope regime delineated clearly by M\,101 outer points; and, b) $\rm\Sigma_{H_{2}}$/$\rm\Sigma_{gas}$\,$\sim$\,0.2, above which the points from both M\,101 and NGC\,628 sample the observed correlation, though NGC\,628 is clearly dominating in this regime. For higher oxygen abundances, up to 12+log(O/H) $\approx$ 8.8, ($\sim$1.4 solar), the negative slope of the relation GDR vs. (O/H) is much flatter than the one for the lower abundance branch, reaching to 12+log(O/H) = 7.9. In contrast, against $\rm\Sigma_{H_{2}}$/$\rm\Sigma_{gas}$, the GDR behaves nearly insensitive to the molecular gas fraction below $\rm\Sigma_{H_{2}}$/$\rm\Sigma_{gas}$\,$\sim$\,0.2, whereas a clear and definite correlation of GDR with $\rm\Sigma_{H_{2}}$/$\rm\Sigma_{gas}$ can be appreciated for the regions located above this value (see Figs.~\ref{fig:GDR_metal} and~\ref{fig:GDR_fracmolgas}). Therefore we propose an empirical relationship between GDR and the combination of these two parameters, 12+log(O/H) for metallicity, and $\rm\Sigma_{H_{2}}$/$\rm\Sigma_{gas}$ which is used as a proxy of the cold molecular clouds relative content, we believe these relations can describe better the complex behaviour of the GDR over the entire range of variations observed here. Empirical fits for these relations have been derived as follows: 

\begin{itemize}
\item[a)] for 12+log(O/H) $\geq$ 8.3 and log(M$\rm_{H_{2}}$/M$_{\rm gas}$) $\geq$ -0.05:
\begin{eqnarray}
\rm log(GDR) = (-0.101\pm0.183)\times (12+log(O/H)) +  \nonumber \\ 
\rm (-0.199\pm0.094)\times log(\Sigma_{H_{2}}/\Sigma_{gas}) + (3.299\pm1.600) \nonumber
\end{eqnarray}

\item[b)] When 12+log(O/H)\,<\,8.3 (i.e. from M\,101 points): 
\begin{eqnarray}
\rm log(GDR) = (-1.332\pm0.048)\times (12+log(O/H)) +  \nonumber \\ 
(13.718\pm0.396)
\end{eqnarray}
\end{itemize}

The relation between GDR and 12+log(O/H) at low metallicity depends on the star formation time scale, showing substantial scatter for a given O/H, as seen in the observations \citep[e.g.][]{2014A&A...563A..31R} and, according to theoretical models, it also depends on the critical metallicity \citep[e.g.][]{2013EP&S...65..213A}. As indicated before for oxygen abundances below 12+log(O/H)\,<\,8.3, Eq.\,4 has been obtained from a fit to M\,101 points only. For M\,101 the critical metallicity inferred here is consistent with the value found for NGC\,628, to within the errors, and similar to the one for M\,33 derived by \citet{2018A&A...613A..43R}, who presented a spatially resolved GDR versus 12+log(O/H) relation in line with our findings. Bearing in mind the small number of low metallicity discs studied so far with spatially resolved data, we would expect comparable values of the critical metallicity to be derived for those galaxies for which M\,101, NGC\,628 and M\,33 can be seen as representative objects (i.e. sharing similar star formation activity, star formation time scales -few Gyr-, chemical abundances and related properties). For these galaxies, the relation between GDR and 12+log(O/H) in the low metallicity range of Eq.\,4 could bring us a guide, though with large uncertainty until more spatially resolved studies are available.

Fig.~\ref{fig:GDR_fracmolgas} illustrates very well the shallow slope found here for the GDR versus molecular fraction relation and also the small role ($\sim$30$\%$ change) played by the oxygen abundance in the fit for the inner galactic regions presenting higher molecular fraction.  

\subsection{Stellar dust production}
\begin{figure*} 
\includegraphics[width=0.47\textwidth]{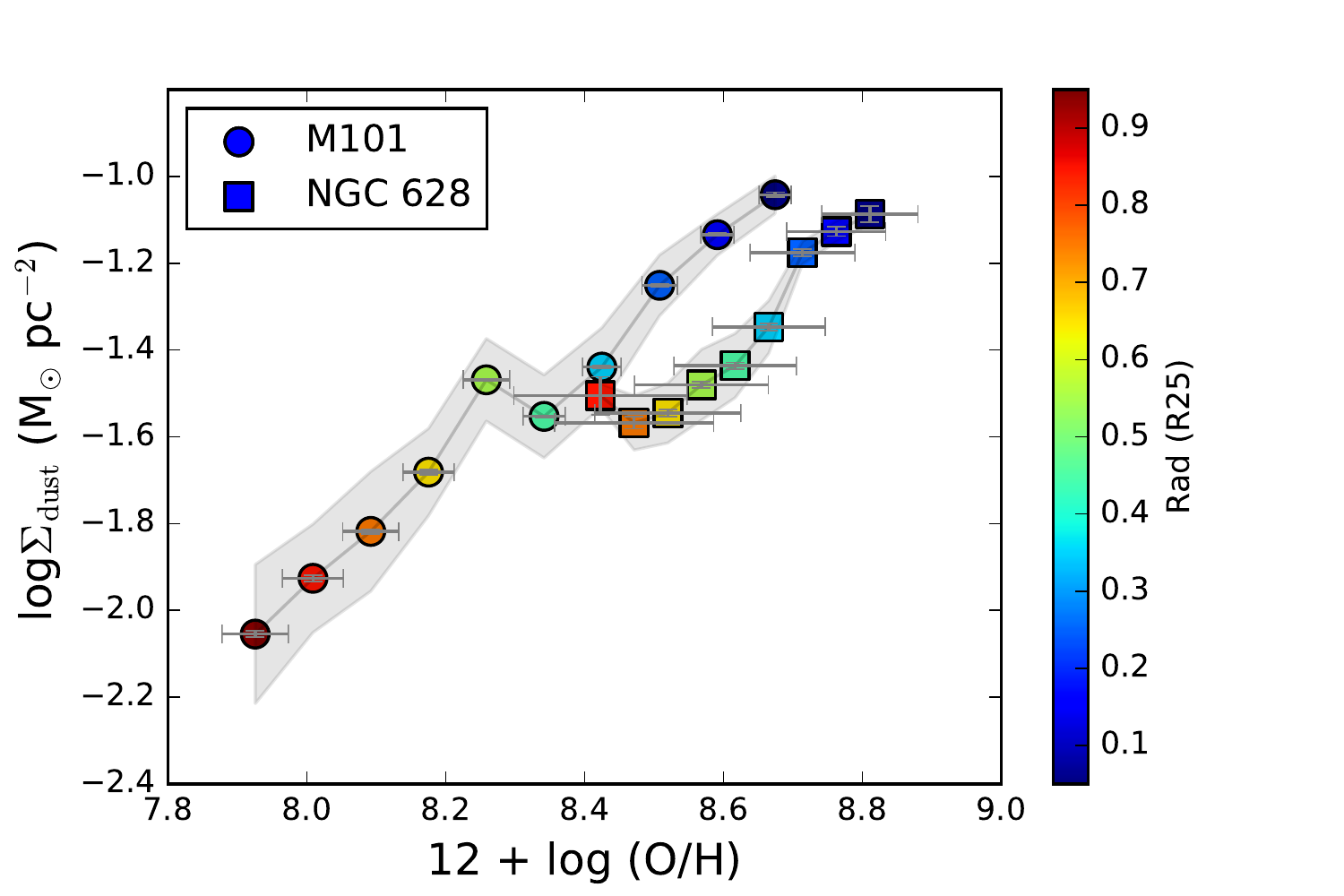}
\includegraphics[width=0.47\textwidth]{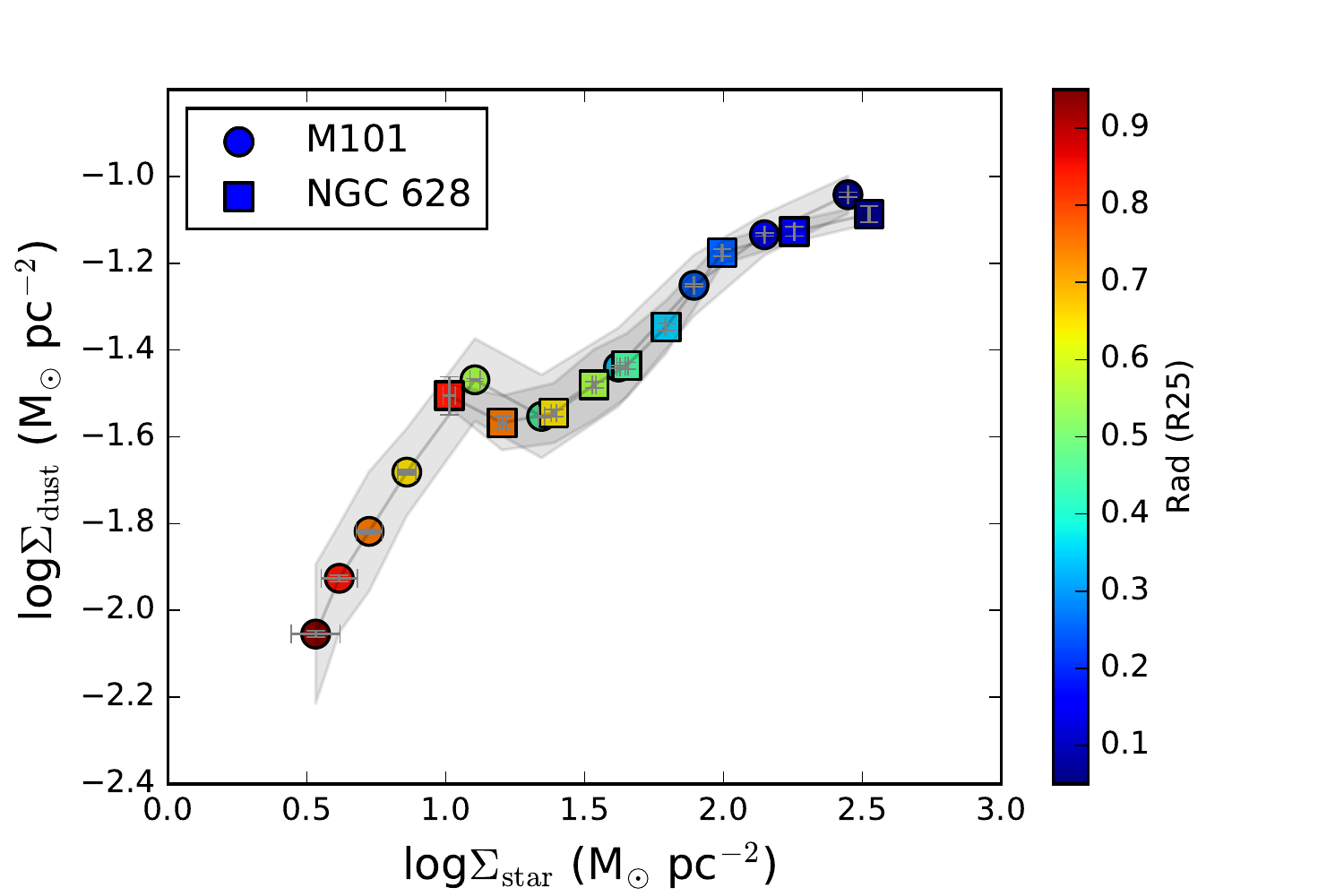}
   \caption{Logarithmic of $\rm\Sigma_{dust}$  versus 12+log(O/H) (left) and versus logarithmic of $\rm\Sigma_{star}$ (right) for M\,101 (circles) and NGC\,628 (squares).}
   \label{fig:SDDust_others}
\end{figure*}

At this stage we can use the predictions from chemical evolution models, including metals and dust production/destruction in galaxies \citep[e.g.][]{2012MNRAS.423...38M,2012MNRAS.423...26M}, in order to try to understand the results obtained. As a matter of fact, we can proceed exploring the component of stellar production of dust, which according to models could be associated to the stellar {\it 'dust yield'}. In this respect, we have searched for a correlation between the spatially resolved dust surface density profile and the oxygen abundance spatial profile for both galaxies. In the left panel of Fig.~\ref{fig:SDDust_others} the relation is presented of the dust mass surface density (in M$_{\odot}$/pc$^{2}$) versus the metallicity as derived across M\,101 and NGC\,628. An overall correlation can be seen with the lowest metallicity points (i.e. outermost regions of M\,101) associated to the lowest dust mass surface density but growing slowly up to 12+log(O/H)\,$\sim$\,8.3 to 8.4, where a discontinuity is apparent, above which the slope clearly increases again, though showing a systematic difference in oxygen abundance of $\sim$0.15 dex higher for NGC\,628 points; possibly a non negligible abundance difference, but here we should bear in mind that for NGC\,628, although the average fit to the oxygen abundance gradient w.r.t. $\rm R/R_{25}$ showed systematically higher abundances than in the case of M\,101, its scatter was also larger ($\sim$$\pm$0.15 dex) than for M\,101.  

In the right panel of Fig.~\ref{fig:SDDust_others} the dust mass surface density is plotted against the stellar mass surface density across both galaxies. A strong and clear correlation is observed and now all the points for both galaxies are located following well the same relation, without an obvious discrepancy in the loci of both galaxies. Again two main regions with two (somewhat) different slopes are present, below and above of a (small) discontinuity apparent for a dust surface density of log\,($\Sigma_{\rm dust}$/(M$_{\odot}$\,pc$^{-2}$))\,$\approx$\,-1.5  (this value corresponding to the region around critical metallicity, as can be read in the left panel), with the steeper slope derived for log\,$\Sigma_{\rm dust}$\,<\,-1.5 (in the same units) corresponding to the regions of lower metallicity and lower stellar density. This result can be expected according to chemical evolution dust models \citep[e.g.][top panels of their Fig. 4]{2012MNRAS.423...26M}, from which we can anticipate that, for metallicities below Z$\approx$0.01, all models covering a large range in dust growth parameter predict that the dust mass produced should be always proportional to the metallicity, for a large range (within a factor of 10$^{3}$) in the dust yield to metal yield ratio; the absolute level of the dust mass produced being proportional to the dust yield adopted. Above Z$\approx$0.01, the role played by the dust growth parameter appears very important, and the relation between the logarithmic of the dust mass and the logarithmic of the metallicity appears non linear, with (at least) two other zones in the relation with different shapes when going to higher (super solar) metallicities, as observed in this work. 

In Fig.~\ref{fig:SDDust_others}, near the discontinuity seen in the right panel mentioned above, a (small) local maximum in the log$\rm\Sigma_{dust}$ versus log$\rm\Sigma_{star}$ relation of M\,101 appears, traced by the points for a value of  $\rm\Sigma_{star}$$\sim$\,10 M$_{\odot}$/pc$^{2}$.
For this local maximum the corresponding normalized radius is $\rm R/R_{25}\sim$\,0.55, equivalent to $\sim$\,7.9\arcminut \ in the disc of M\,101. It is worth noting here that M\,101 presents a well behaved rotation curve in the inner part of the galaxy for radial distances lower than 7\arcminut , whereas a distorted outer part has been revealed for radii \,>\,7\arcminut \ \citep{1981A&A....93..106B}. The radial distance of the observed local maximum of $\sim$\,7.9\arcminut \ appears just beyond the limit defining the outer distortet part of M\,101. Interestingly enough, over the radial range 7 - 9\arcminut , \citet{2013ApJ...762...82M} have found changes in the photometric profile slope of M\,101 for the various octants of the disc analysed.

\begin{figure} 
\includegraphics[width=0.47\textwidth]{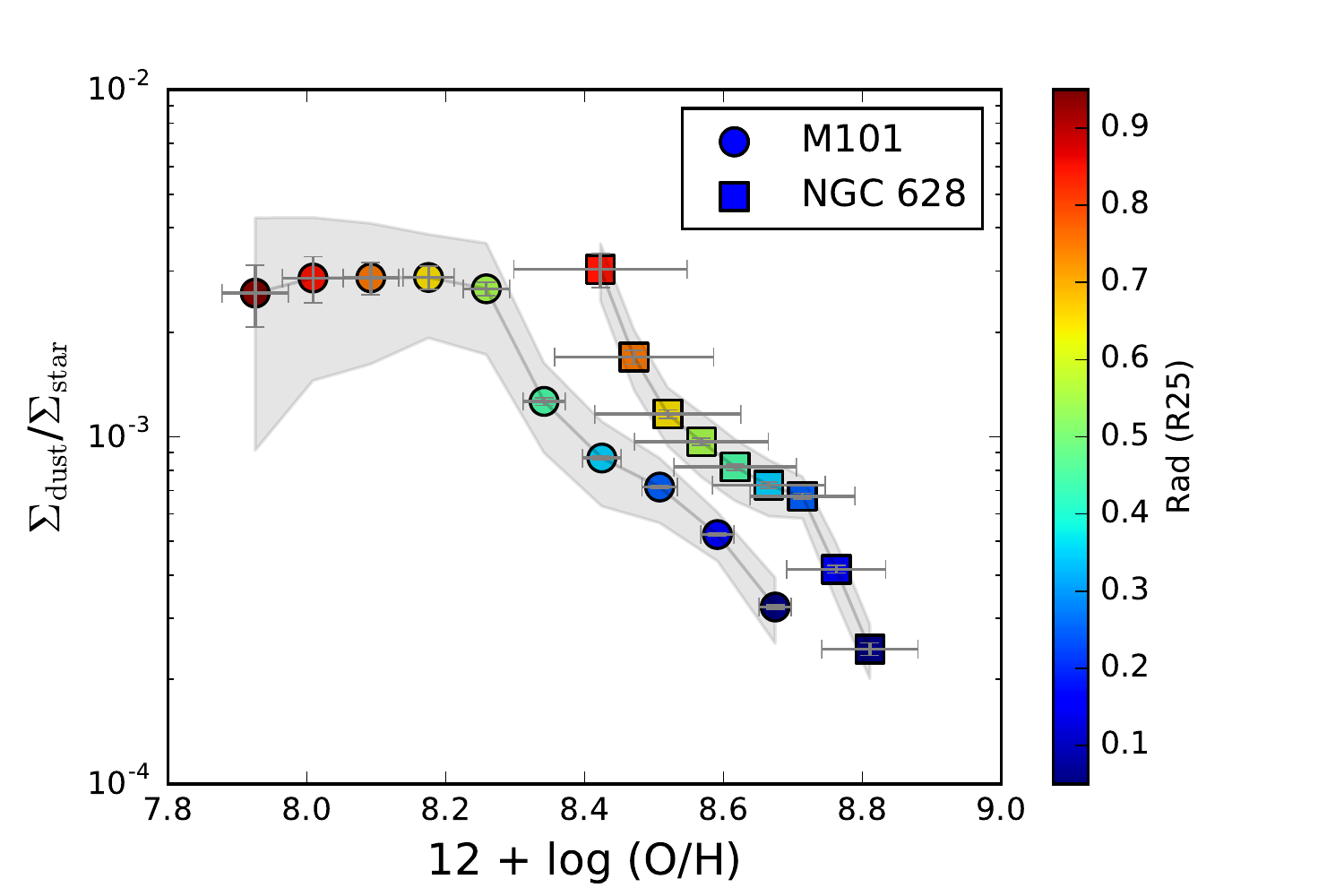}
\includegraphics[width=0.47\textwidth]{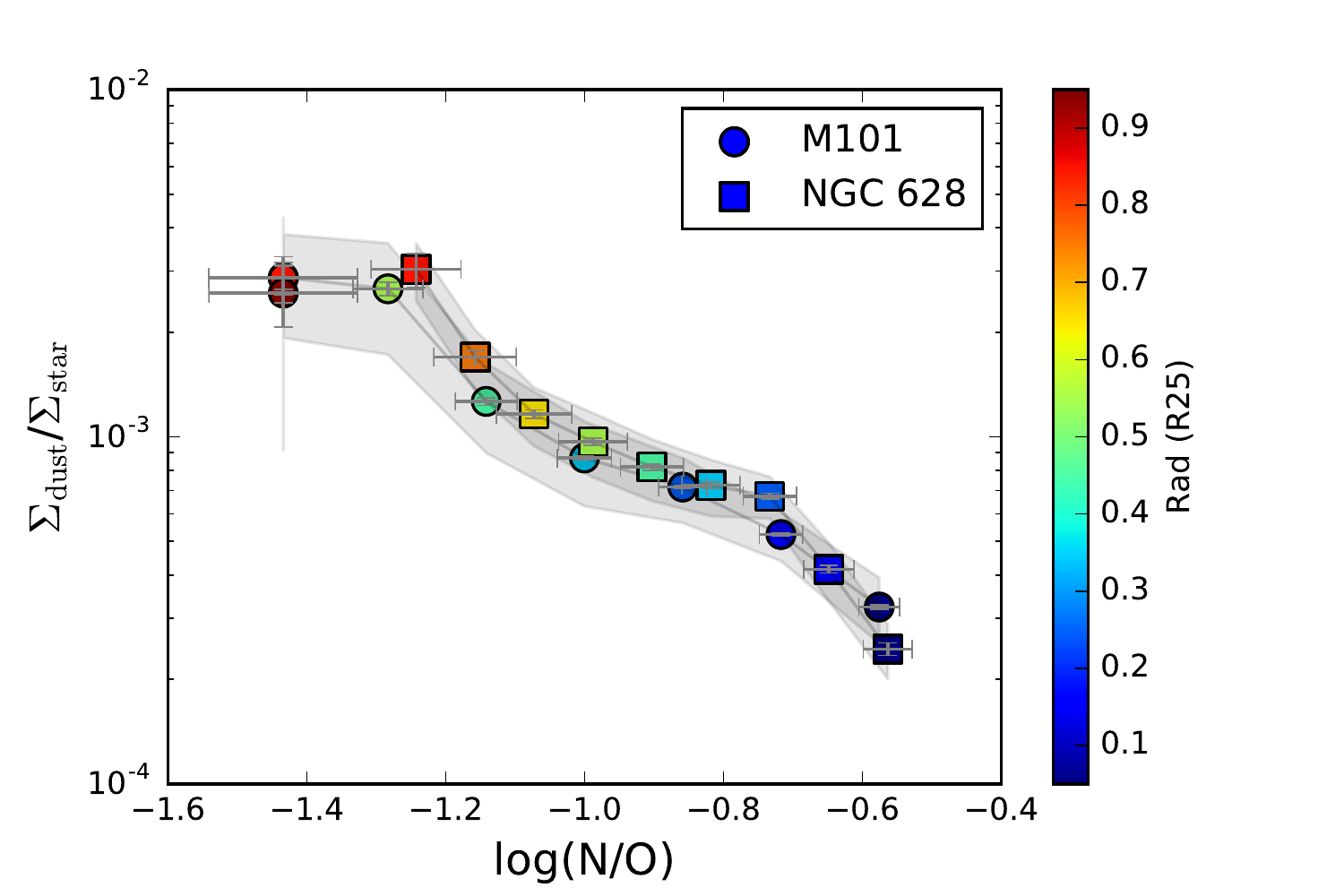}
\includegraphics[width=0.47\textwidth]{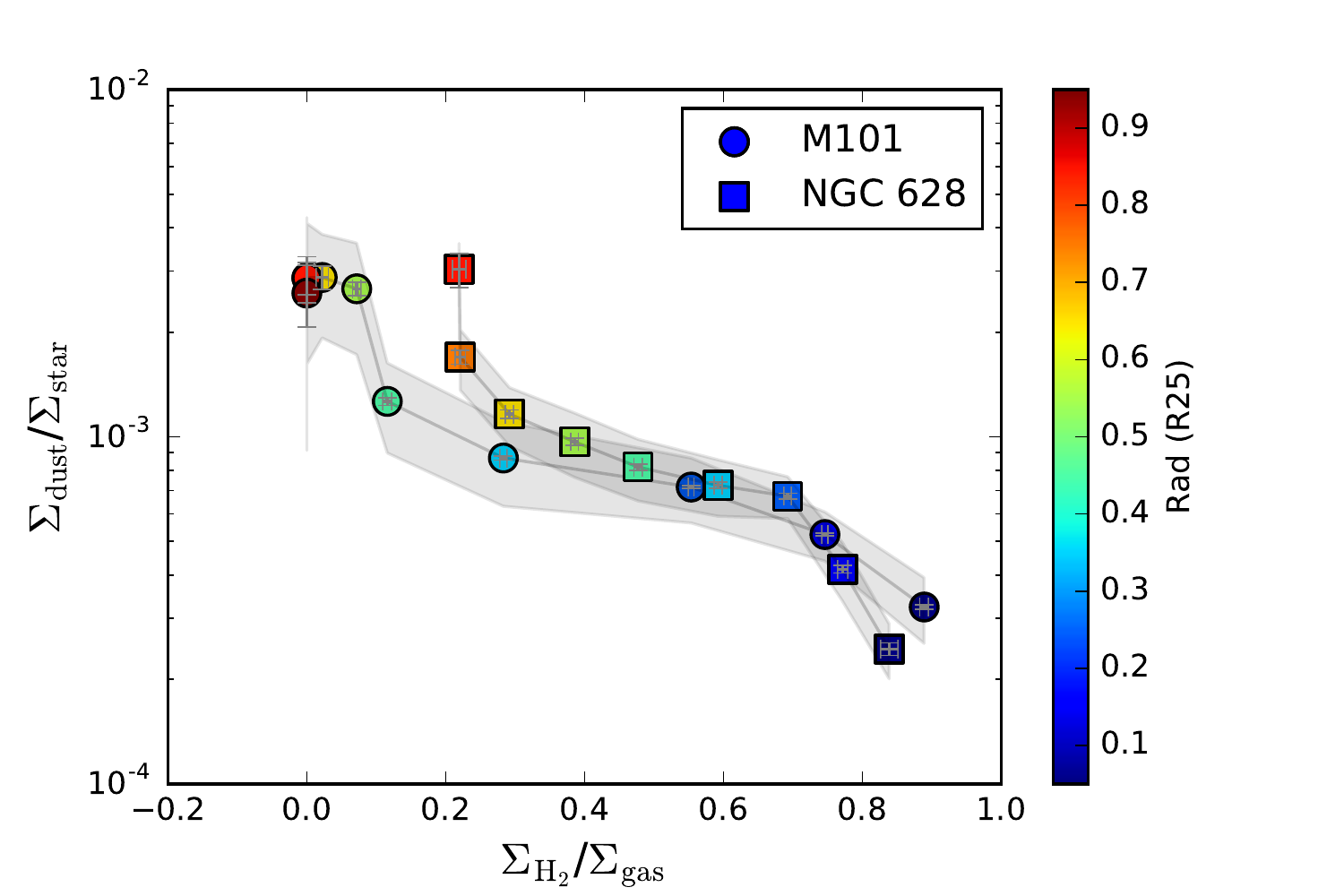}
   \caption{$\rm\Sigma_{dust}$ over $\rm\Sigma_{star}$ versus 12 + log(O/H) (top), log(N/O) (middle) and molecular gas mass fraction (bottom) for M\,101 (circles) and NGC\,628 (squares).}
   \label{fig:MdustMstar}
\end{figure}

We have explored further the role of $\rm\Sigma_{dust}$/$\rm\Sigma_{star}$ as a proxy for an empirical estimation of the stellar dust yield ({\it stardust}). $\rm\Sigma_{dust}$ is a balance between the amount of dust produced by the stars, that one destroyed by SN shocks and the amount formed by accretion in molecular clouds. In the top panel of Fig.~\ref{fig:MdustMstar} we show $\rm\Sigma_{dust}$/$\rm\Sigma_{star}$ versus oxygen abundance 12+log(O/H). The relation observed presents a constant -within the errors- value of $\rm\Sigma_{dust}$/$\rm\Sigma_{star}$ $\approx$0.003 (close to the value of the oxygen yield derived here, see Sec.\,\ref{sec:chemevol}) corresponding to low metallicity in the outer galaxy points. We expect that, as there is no molecular gas at these outer radii, the contribution of accretion to the dust budget is minimal, and therefore the amount of dust in this region is related to stellar dust production and destruction. This situation is holding until the critical abundance for M\,101 is reached. Above 12+log(O/H) $\geq$ 8.3 to 8.4 a strong correlation is seen for both galaxies with a negative slope decreasing the empirical stardust production by an order of magnitude to $\rm\Sigma_{dust}$/$\rm\Sigma_{star}$ $\approx$  0.0003 for supersolar abundance, 12+log(O/H) $\approx$ 8.8. Again, above this value the oxygen abundances appear to be slightly larger (by $\sim$ 0.15 dex) for NGC\,628 than for M\,101.

It is clear that the ratio $\rm\Sigma_{dust}$/$\rm\Sigma_{star}$
should depend on the star formation history of the system, spatially resolved across the discs of the galaxies. Since the absolute value of the total stellar mass across these galaxies can certainly be weighting the derivation of this ratio, we also have presented it against the N/O abundance ratio (see below), an intensive parameter which can be seen as a 'chemical clock' tracing the main nitrogen star producers, of typical age of order several Gyr, compared to $\sim$1 Gyr for the assumed dust life time \citep[][]{2017arXiv171107434G}. We note here that the data available do not allow to perform a detailed study in order to differentiate the two main contributors to the stellar dust: AGB stars, that inject dust in a timescale of 100-500\,Myr depending on the star formation history, IMF and metallicity \citep{2009MNRAS.397.1661V}; and SNe producing dust a few hundred years after the explosion \citep{2001MNRAS.325..726T,2007MNRAS.378..973B}. The stellar population studies available in the literature show radial age profiles that vary between  1-10\,Gyr for NGC\,628 \citep{2011AJ....142...16Z,2014MNRAS.437.1534S} and 2-7\,Gyr for M101 \citep{2013ApJ...769..127L}.

In the middle panel of Fig.~\ref{fig:MdustMstar}, $\rm\Sigma_{dust}$/$\rm\Sigma_{star}$ is now shown against the nitrogen to oxygen ratio, log(N/O), a proxy of the secondary nucleosynthesis production and thus an empirical  'chemical clock'. The correlation is strong now with no apparent difference in the relation between the points from both galaxies: all points sharing now the same locus on the plot. Therefore, we can say that N/O behaves as a good tracer of dust in this range -possibly better than the oxygen abundance-. We can see how for the abundance range of primary nitrogen production, i.e. represented here by the assumed log(N/O) $\approx$ -1.4, all (M\,101) points show a value of $\rm\Sigma_{dust}$/$\rm\Sigma_{star}$ constant and equal to the value quoted from the previous plot. Then, for larger values of log(N/O) across both galaxies -i.e. going towards the inner parts of the disc where gas is expected to be more chemically evolved with a secondary nitrogen production- points show a very low $\rm\Sigma_{dust}$/$\rm\Sigma_{star}$ ratio. We speculate if this may suggest possible dust destruction mechanisms operating in addition to the ISM dust growth process mentioned before. However, it seems plausible that for the inner regions of these galaxies the $\rm\Sigma_{dust}$/$\rm\Sigma_{star}$ ratio may be lower as a consequence of their higher past star formation, leading there to their higher stellar mass and faster chemical enrichment as inferred from the available data.

According to chemical evolution models with dust production \citep[e.g.][]{2012MNRAS.423...26M} the chemical evolution of the 'stardust' production could be described as the one for an element with primary and secondary nucleosynthesis, as it is the case for nitrogen. In this case and following the theoretical prescriptions, our simple description would indicate that the linear part at low N/O should be tracing directly primary 'stardust' production, whereas the secondary part would be associated to a metallicity dependent dust yield, that assuming no growth nor destruction of dust in the ISM. In this scheme, for the low (primary) metallicity regime, the quotient between the dust to gas ratio and the metallicity would be equal to the ratio of the primary dust yield and the metal yield. In that case, our results above favour a value of the dust yield similar to the oxygen yield derived in this work, i.e. some 40$\%$ of the total metals yield \citep[e.g.][]{2012MNRAS.423...38M}. Of course, since we here directly compared dust mass to stellar mass, we are dealing with the 'stardust' yield part of the dust production. Clearly, the corresponding dust production associated to ISM dust growth need to be added in further work.
 
In the bottom panel of Fig.~\ref{fig:MdustMstar} we have explored the behaviour of $\rm\Sigma_{dust}$/$\rm\Sigma_{star}$ versus the surface density of the molecular cold gas fraction, $\rm\Sigma_{H_{2}}$/$\rm\Sigma_{gas}$, as our empirical indicator of the galactic regions of ISM dust growth production predominance. We can see that the slope is again negative but now the range with a much shallower slope seems large, presenting a nearly similar value of $\rm\Sigma_{dust}$/$\rm\Sigma_{star}$ for a substantial fraction of $\rm\Sigma_{H_{2}}$/$\rm\Sigma_{gas}$ from $\approx$0.2 up to 0.7, for both galaxies. The highest changes in $\rm\Sigma_{dust}$/$\rm\Sigma_{star}$ in this plot are seen at the two extremes of very low (high 'stardust' production) and very high (low 'stardust' production) $\rm\Sigma_{H_{2}}$/$\rm\Sigma_{gas}$. 

Finally, it is worth mentioning that our interpretation of the empirical trends presented in this paper proposes a change of the main mechanisms contributing to the dust mass budget, from accretion at high oxygen abundances, to stellar dust production for oxygen abundance values lower than 12+log(O/H)$\sim$8.4. However, we should keep in mind that this value of a 'critical metallicity' would agree with the metallicity at which the relative fraction to PAHs is expected to decrease from $\sim$3\% to values less than 1\% at lower metallicities \citep{2007ApJ...663..866D}. Further future work to study possible physical connections between the observed trends of GDR and the relative abundance of PAHs would be interesting. 

\section{Conclusions}\label{sec:conclusions}

In this work we have studied in detail the relation between dust and metallicity in the nearly face on nearby galaxies NGC\,5457 (M\,101) and NGC\,628 (M\,74) using high quality data available (including optical spectroscopy of the ionised gas, multi-band spectrophotometry of the stars, dust and gas, radio and millimetre observations of their neutral and molecular gas). Both galaxies can be considered prototype objects of its class and have a similar mass (log\,M$_{\star}$/M$_{\odot}\approx$\,10.2), though they present apparently different environments and chemical and structural properties. We have combined the chemical abundances of the ISM and the stars of both galaxies with a detailed derivation of their dust content and of their chemical evolution. The main results obtained in this work are the following:

\begin{itemize}
\item The global budgets of metals and dust, and the chemical evolution, across M\,101 and NGC\,628 have been derived and analysed consistently for both galaxies, under the same physical assumptions. We have found that for both galaxies the overall metal budget could be consistent with the predictions of the simple model of chemical evolution if we assume an oxygen yield around half solar to one solar, close to the observational derivations (y$\approx$0.003) and to some recent theoretical prescriptions, though other nucleosynthesis schemes predict higher values. Whereas the whole disc of NGC\,628 appears to be consistent with this result, M\,101 presents a deviation in the outermost region (R\,$\geq$\,0.8\,$\rm R_{25}$), suggesting non closed box gas flows there. 

\item The gas to dust mass ratio map for each galaxy is obtained from the maps of total gas mass and the dust mass that have been computed. Despite the presence of sufficient small scale structure across the maps, NGC\,628 GDR seems to be relatively constant across the disc of the galaxy, while the GDR of M\,101 shows a significant increase towards the outer parts of the galaxy. The GDR radial profiles of the galaxies show different behaviour with galactocentric radii, with NGC\,628 increasing very smoothly with galactocentric radius, while for M\,101 the GDR shows a significant increase towards the outer parts of the galaxy.

\item The gas to dust mass ratio varies against the chemical abundance obtained across the galaxies showing a two slopes behaviour, with a break at 12+log(O/H)\,$\approx$\,8.4; showing a steep negative slope ($\Delta$ logGDR/$\Delta$ log(O/H)$\approx$ -1.3) below this oxygen abundance. This result is found to be consistent with theoretical predictions which define a critical metallicity for this slope change \citep[e.g.][]{2013EP&S...65..213A}, in the line of what is seen in recent observational studies for M\,33 \citep[][]{2018A&A...613A..43R} and from previous samples of star forming objects \citep[e.g.][]{2013ApJ...777....5S,2014A&A...563A..31R}. 

\item We have found an empirical relation between the gas to dust ratio, metallicity and the fraction of molecular to total gas mass ratio, $\rm\Sigma_{H_{2}}$/$\rm\Sigma_{gas}$. The GDR shows a very close relationship with metallicity for the lowest abundances, and with $\rm\Sigma_{H_{2}}$/$\rm\Sigma_{gas}$ for the higher metallicity, leading to the lower GDR values (GDR\,$\lesssim$\,400) measured, suggestive of ISM dust growth.

\item It has been found that log\,$\rm\Sigma_{dust}$, the dust surface density, is well correlated with 12+log(O/H) and also with log\,$\rm\Sigma_{star}$, the stellar surface density, in agreement with theoretical chemical evolution models including dust \citep[e.g.][]{2012MNRAS.423...38M,2012MNRAS.423...26M}. In the same vein, we have found that across both galaxies the ratio $\rm\Sigma_{dust}$/$\rm\Sigma_{star}$, an empirical proxy for the stellar dust production, is well correlated with oxygen abundance, 12 + log(O/H), and appears strongly correlated with log(N/O), i.e. the nitrogen to oxygen abundance ratio which is a direct indicator of primary to secondary nucleosynthesis, and which we propose as a tracer of the GDR and of the 'stardust' production behaviour. For abundances below the critical abundance, the 'stardust' production as measured by $\rm\Sigma_{dust}$/$\rm\Sigma_{star}$, gives a constant value (mainly in the M\,101 outer regions) indicating that the (stellar) dust yield is similar to the value of the oxygen yield derived for the region (equivalent to some 40$\%$ of the total metal yield). 

\item We have found a relation of $\rm\Sigma_{dust}$/$\rm\Sigma_{star}$ with the molecular gas mass fraction, $\rm\Sigma_{H_{2}}$/$\rm\Sigma_{gas}$, which could illustrate the balance between 'stardust' production and ISM dust growth for M\,101 and NGC\,628. This relation shows a complex behaviour with a nearly constant 'stardust' component for a large range in molecular gas mass fraction values (0.2\,$< \rm\Sigma_{H_{2}}/\rm\Sigma_{gas} \leq$\,0.8); being the largest changes in $\rm\Sigma_{dust}$/$\rm\Sigma_{star}$ seen at the extremes of low and high $\rm\Sigma_{H_{2}}$/$\rm\Sigma_{gas}$. More deep observations and detailed analyses are needed before going further in this direction. A more quantitative analysis of the empirical scenarios here presented is underway, including data of three nearby spirals and the Milky Way. We are presently developing the application of the 'stardust' low-metallicity scenario with no dust growth, illustrated in this work, to a sample of low metallicity dwarfs to be presented in a forthcoming work (Rela\~no et al. in preparation).
\end{itemize}

\section*{Acknowledgements}
The authors would lilke to thank the anonymous referee for very constructive comments that have helped to improve the paper. This work was partially supported by the Spanish Ministerio de Econom\'{\i}a y Competitividad under grants AYA2016-79724-C4-4-P and AYA2016-79724-C4-3-P, and excellence project PEX2011-FQM-7058 of Junta de Andaluc\'{\i}a (Spain).  MR acknowledges support by the research projects AYA2014-53506-P and AYA2017-84897-P from the Spanish Ministerio de Econom\'{\i}a y Competitividad and Junta de Andaluc\'{\i}a grant FQM108; support from the European Regional Development Funds (FEDER) is acknowledged. JMV thanks the Director of the IoA for hospitality, and the Spanish MECD for a "Salvador de Madariaga" grant PRX17/00485, and the MIAPP 2016 program "The Chemical Evolution of Galaxies" of the Excellence Cluster "Universe" for partial support. 
We thank Miguel Querejeta for providing the stellar mass maps of M\,101 and NGC\,628, Karin Sandstrom for the \hi\ and CO data, and the KINGFISH collaboration for providing us with \Spi\ and \Her\ data of the galaxy sample. Thanks are given also to R. Asano for providing the evolutionary tracks of his models.
MR thanks the Computational service PROTEUS at the Instituto Carlos I (Universidad de Granada). This research made use of APLpy and Matplotlib an open-source plotting package for Python.




\bibliographystyle{mnras}
\bibliography{jvm}

\bsp	
\label{lastpage}
\end{document}